\documentclass[a4paper,11pt]{article}
\usepackage[DIV12]{typearea}
\usepackage{longtable}
\usepackage{booktabs,colortbl}
\usepackage{multirow}
\usepackage{amsmath}
\usepackage{amssymb}
\usepackage{bbm}
\usepackage[utf8]{inputenc}
\usepackage{pdflscape}
\usepackage{xspace}
\usepackage[caption=false]{subfig}
\usepackage{nicefrac}
\usepackage{graphicx}
\usepackage{cite}  
\usepackage[small]{caption2}
\usepackage{mathtools}
\usepackage{nccfoots}
\usepackage{tikz}

\usetikzlibrary{calc,positioning}

\usepackage{pgfplots}%
\pgfplotsset{compat = newest}

\usepackage[all,cmtip]{xy}
\newlength{\xywd}
\newcommand{\xyrightarrow}[2][]{%
  \sbox{0}{$\scriptstyle#1$}%
  \xywd=\wd0
  \sbox{0}{$\scriptstyle#2$}%
  \ifdim\wd0>\xywd \xywd=\wd0 \fi
  \xymatrix@C\dimexpr\xywd+1em\relax{{}\ar[r]^{#2}_{#1}&{}}%
}

\newcommand{\Z}[1]{\ensuremath{\mathbbm{Z}_{#1}}} % z_N ->\Z{N}

\newcommand{\U}[1]{\ensuremath{\mathrm{U}(#1)}}

\newcommand{\I}{\mathrm{i}}

\newcommand{\CP}{\ensuremath{\mathcal{CP}}\xspace}
\newcommand{\x}{\ensuremath{\times}}

\usepackage{xcolor}
\definecolor{darkgreen}{HTML}{109930}
\definecolor{pink}{rgb}{0.858, 0.188, 0.478}

\advance \headheight by 3.0truept       % for 12pt mandatory...
\usepackage[pdftex]{hyperref}
\hypersetup{
    pdftitle = {The simplest case of dark inflaxion},
    pdfauthor = {Gordillo-Ruiz, Morales-Mena, Ramos-Sanchez}
}

\begin{document}

\begin{titlepage}

%\vspace*{-3.0cm}

\vspace*{1.0cm}

\begin{center}
{\Large\textbf{\boldmath 
The simplest case of dark inflaxion}
\unboldmath}

\vspace{1cm}
\textbf{Hansel Gordillo--Ruiz},
\textbf{C\'esar Morales--Mena}, and
\textbf{Sa\'ul Ramos--S\'anchez}
\Footnote{*}{%
\href{mailto:hanselgordillo@ciencias.unam.mx;cesarfaim@comunidad.unam.mx;ramos@fisica.unam.mx}{\tt hanselgordillo@ciencias.unam.mx;cesarfaim@comunidad.unam.mx;ramos@fisica.unam.mx} 
}
\\[5mm]
\textit{\small Instituto de F\'isica, Universidad Nacional Aut\'onoma de M\'exico,\\ POB 20-364, Cd.Mx. 01000, M\'exico}
\end{center}

\vspace{1cm}

\vspace*{1.0cm}

\begin{abstract}
Dark matter and cosmic inflation represent two of the major puzzles in cosmology. 
They are typically addressed by introducing separate fields with independent dynamics.
On the other hand, extra dimensions might play an important role for observable
physics. We introduce a five-dimensional model called {\it dark inflaxion} that includes an axion-like particle,
whose Kaluza-Klein modes can describe inflation and dark matter. We show
that a simple yet natural choice of the mass scale of the effective four-dimensional 
fields of our model can accommodate simultaneously the observable values of inflationary
parameters and dark-matter abundance. It will be interesting to explore the
consequences and predictions of more generic scenarios within the scope of our model,
which include the possibility of multifield inflation and dark matter.
\end{abstract}

\end{titlepage}

\newpage

%%%%%%%%%%%%%%%%%%%%%%%%%%%%%%%%%%%%%%%%%%%%%%%%%%%%%%%%%%%%%%%%%%%%%%%%%%%%%%%%%%%%%%%%%%%%%%%%%%%%%%%%%%%%%%

\section{Introduction}  

The $\Lambda$CDM standard model of cosmology is the best available fit of the increasingly
precise cosmological measurements, which include temperature fluctuations of the 
cosmic microwave background (CMB)~\cite{Hu:2001bc,Planck:2018nkj}. This description 
indicates that our Universe experienced an early era of exponential expansion known 
as cosmic inflation~\cite{Guth:1980zm,Linde:1981mu}, which may have lasted less 
than $10^{-30}\,$s~\cite{Kolb:1990vq}, and which is responsible for many of the observed
features of our Universe, including its apparent flatness~\cite{Liddle:1999mq,Tsujikawa:2003jp,Vazquez:2018qdg,Achucarro:2022qrl}.
Another consequence of the model is that about 85\% of the matter around us must be dark 
matter (DM), although its precise nature is unknown.

There are plenty of inflationary models consistent with data (see e.g.~\cite{Martin:2013tda,Martin:2013nzq}). 
In such models, the dynamics of a scalar or pseudo-scalar field $\phi$ called {\it inflaton} 
with proper state equation is responsible for cosmic inflation. In this work, we are interested in 
cosmological features arising from axion-like particles (ALPs), which are pseudo-scalar gauge 
singlets, similar to the axions introduced to solve the strong \CP 
problem~\cite{Peccei:1977hh,Wilczek:1977pj,Weinberg:1977ma} (see e.g.~\cite{Marsh:2017hbv} 
to learn the basics of ALPs) with rich cosmological phenomenology~\cite{Marsh:2015xka,Sikivie:2006ni,Hu:2000ke}.
Further, it results interesting that they appear naturally in top-down constructions~\cite{Svrcek:2006yi,Arvanitaki:2009fg}. 
Some examples of successful inflationary models based on ALPs include natural 
inflation~\cite{Adams:1992bn,Freese:1990rb} and its extensions~\cite{Kim:2004rp,Czerny:2014wza},
$N$-flation~\cite{Dimopoulos:2005ac}, hilltop inflation~\cite{Boubekeur:2005zm,Takahashi:2021tff},
models of axion monodromy~\cite{Silverstein:2008sg,McAllister:2008hb,Kappl:2014lra} and other
string-inspired scenarios~\cite{Dvali:2001fw,Burgess:2008ir,Silverstein:2003hf,Banks:1995dp,Green:2009ds,Damian:2018tlf,Kachru:2003sx,Garcia-Bellido:2001lbk,Cicoli:2008gp, Conlon:2005jm,Avgoustidis:2006zp,Cicoli:2011ct,Cicoli:2023opf},
among others.

On the other hand, there exist also many models that address the indirect evidence 
of the existence of DM, mostly based on CMB~\cite{Planck:2018vyg}. In most of the models, 
one postulates additional gauge neutral particle(s) whose various properties lead to 
compatibility with observations, complying also with bounds based on direct-detection 
searches~\cite{XENON:2018voc} (see~\cite{Arbey:2021gdg} and references therein). Besides
the traditional WIMP paradigm~\cite{Roszkowski:2017nbc}, it has been shown that DM can 
very well be composed of (typically ultralight) ALPs~\cite{Matos:1999et,Hu:2000ke} or
some Kaluza-Klein (KK) modes of fields in five-dimensional (5-d) 
scenarios~\cite{Dienes:2016zfr,Cheng:2002ej,Servant:2002aq}. 
An interesting possibility of this kind is the so-called dynamical 
DM (DDM)~\cite{Dienes:2011ja,Dienes:2011sa,Dienes:1999gw}. In this scenario, DM is obtained from
the infinite KK-tower states of an ALP in extra dimensions; and it is dynamic
because these DM components decay, yet there is a fine balance between their abundances
and decay widths, leading to the observed DM abundance today $\Omega_\mathrm{DM}$.
This scheme partially inspires our framework.

In the quest for a minimalistic model of cosmology, it is appealing to conceive
a construction in which a reduced number of elements accounts for the answers
to two or more open questions. In this context, ALPs seem to be instrumental to
arrive at a simultaneous plausible explanation of the nature of DM and 
inflation~\cite{Daido:2016tsj,Daido:2017wwb,Daido:2017tbr,Cheng:2021qmc}. Our 
work adds to this pursuit. We introduce a new model which we call {\it 
dark inflaxion}, in which a 5-d ALP provides the Universe with the virtues of
dark matter and inflation.
The resulting effective scenario is similar in some aspects to the
ALP miracle~\cite{Daido:2017wwb,Daido:2017tbr}, even though the axion decay
constant is forced to be large, as happens most naturally especially in top-down
models~\cite{Kallosh:1995hi,Banks:2003sx}.

This work is organized as follows. In order to introduce our notation, in 
section~\ref{sec:inflationalps} we present the main features of inflation and ALPs. 
Then, in section~\ref{sec:ALPXD} we introduce the general features of our model based 
on a 5-d ALP and the properties of its KK modes. This leads us to study in section~\ref{sec:benchmark}
the properties of some special cases, including our main example with two undecoupled 
KK modes. In section~\ref{sec:dynamics} we explore 
the dynamics of these modes in a FRW spacetime. We proceed to discuss the 
consequences of the model for inflation in section~\ref{sec:Inflation} 
and for DM in section~\ref{sec:densities}. Our conclusions and outlook are provided
in section~\ref{sec:conclusions}. We devote our appendix to some
side remarks on the viability of small axion decay constants.

\section{Inflation and axion-like particles}
\label{sec:inflationalps}

\subsection{Cosmic inflation}
\label{sebsec:inflation}

Inflation corresponds to an era in which the comoving Hubble length decreases, commonly driven by 
the dynamics of the inflaton $\phi$. Inflationary models with a single inflaton require its potential
$V(\phi)$ to be extremely flat. This guarantees sufficient inflation to comply with the measurements 
of the CMB anisotropies, among other requirements.
The flatness condition can be encoded in the relationship between the height of the potential
and its width~\cite{Adams:1990pn}, 
\begin{equation}
    \frac{\Delta V}{\left(\Delta \phi\right)^4} \leq \mathcal{O}\left(10^{-6}- 10^{-8}\right)\, , 
\end{equation}
where $\Delta \phi$ and $\Delta V$ denote the changes of the inflaton and its potential during 
inflation, respectively.
Solving some issues of the $\Lambda$CDM, such as the horizon and spatial flatness
 problems, requires that the scale factor $\mathrm{a}(t)$, which encodes the size 
of the observed horizon, grows about $10^{26}$ times in less than $10^{-30}$\,s. 
The amount of inflation is measured by the e-folding number,
\begin{equation}
\label{eq:e-folds}
N ~:=~\ln \left( \frac{\mathrm{a}_{\text{end}}}{\mathrm{a}_\text{ini}} \right ) ~=~
 \int_{t_\text{ini}}^{t_{\text{end}}} dt  H  ~\simeq~ 
 \frac{1}{M_{\text{Pl}}^2} \int_{\phi_{\text{end}}}^{\phi_{\text{ini}}}d \phi\,\frac{V }{V'} \, ,
\end{equation}
where $\phi_\text{ini}$ ($\phi_\text{end}$) denotes the field value at the beginning (end)
of inflation, $M_{\text{Pl}}$ is the reduced Planck mass, $H:=\dot{\mathrm{a}}/\mathrm{a}$ denotes
the Hubble parameter, and the primes refer to derivatives of the potential w.r.t. $\phi$. We 
assume $\phi>0$ during inflation without loss of generality. 

For an exponential growth of $\mathrm{a}(t)$ to happen, inflation must take place in a regime 
where $\ddot{\phi}$ is negligible and the Friedmann and continuity equations are approximated as
\begin{equation}
\label{eq:slow-roll_approx}
H^2 ~\simeq~  V(\phi)/3 M^2_{\text{Pl}}\,,
\qquad
3 H\dot{\phi} ~\simeq~ -V'(\phi)\,.
\end{equation}
This is the so-called {\it slow-roll approximation}. In this approximation, it is useful
to characterize the inflationary properties with the help of the (first-order) slow-roll parameters
\begin{equation}\label{eq:slow-roll}
\epsilon ~:=~ \frac{M_\text{Pl}^2}{2} \left(\frac{V'}{V}\right)^2\,, 
\qquad 
\eta ~:=~ M_\text{Pl}^2 \frac{V''}{V}\,. 
\end{equation}
We can also define the second-order parameter
\begin{equation}\label{eq:xi}
\xi ~:=~ M_\text{Pl}^4 \frac{V' V'''}{V^2}   \, .  
\end{equation}
The field $\phi$ could drive inflation as long as $\epsilon, |\eta| < 1$ for enough number of 
e-folds, which is typically taken to be $N\sim60$ or so. We consider that cosmic inflation 
ends when $\epsilon \approx 1$. 

Inflation leaves an imprint in the power spectra of the CMB. In particular, in terms of the 
values of the slow-roll parameters at the beginning of inflation for a certain pivot momentum 
scale $k$, the spectral scalar index is approximated by
\begin{equation}\label{eq:n_s}
    n_s ~\simeq~ 1 + 2\eta - 6\epsilon \, , 
\end{equation}
and the tensor-to-scalar ratio is
\begin{equation}\label{eq:r}
r ~\simeq~ 16\epsilon\, . 
\end{equation}
According to the latest results by Planck and BICEP/Keck 2018, 
the observed values of the scalar spectral index and the tensor-to-scalar ratio are
\begin{subequations}\label{eq:Plancknsrvalues}
\begin{align}
    n_s &~=~ 0.9665\pm 0.0038,        &&\text{({\small TT,TE,EE+lowE+lensing+BAO,~\cite[Table 2]{Planck:2018vyg}})}\, , \label{eq:n_s} \\
    r   &~=~ 0.014^{+0.010}_{-0.011}, &&\text{({\small BK18,~\cite{BICEP:2021xfz}})} \label{eq:r} \, .
\end{align}
\end{subequations}
Both values are evaluated at a pivot scale $k = 0.05\;\text{Mpc}^{-1}$ at a confidence level of $68\%$.
Also the running of the spectral scalar index can be expressed in terms of the
slow-roll parameters~\eqref{eq:slow-roll} and~\eqref{eq:xi},
\begin{equation}\label{eq:running}
\alpha_s ~:=~ \frac{\text{d} n_s}{\text{d}\ln k} ~\simeq~
-2\xi +16\epsilon\eta-24\epsilon^2  \, .   
\end{equation}
Similarly, the amplitude of scalar perturbations is given by 
\begin{equation}\label{eq:As}
    A_s ~=~ \frac{1}{24\pi^2M_\text{Pl}^4}\frac{V}{\epsilon} \, .
\end{equation}
The values of the running of the spectral scalar index and the amplitude of
scalar perturbations reported by Planck are
\begin{subequations}\label{eq:PlanckAsdnsvalues}
\begin{align}
\alpha_s     &= -0.006\pm 0.013 &&({\small 95\%\text{, TT,TE,EE+lowE+lensing+BK15+BAO,~\cite[Eq.~(45b)]{Planck:2018vyg}}})\, , \label{eq:dnsdlnk}\\
    10^9 A_s &= 2.105 \pm 0.030 &&({\small 68\%\text{, TT,TE,EE+lowE+lensing+ BAO,~\cite[Table 2]{Planck:2018vyg}}})\, . \label{eq:As} 
\end{align}
\end{subequations}
Given an inflationary scheme defined by a potential $V$, the observations~\eqref{eq:Plancknsrvalues}
and~\eqref{eq:PlanckAsdnsvalues} serve to constrain the free parameters of the model, as we do here.
Although controversially, to further constrain inflationary models the so-called Lyth bound~\cite{Lyth:1996im},
\begin{equation}
\label{eq:generalLyth}
 \frac{\Delta\phi}{M_\text{Pl}}\geq 2\times \left(\frac{r}{0.01} \right)^{\nicefrac12}\,,
\end{equation}
is invoked as a means to comply with the absence of observations of primordial gravitational
waves in the CMB. Inflationary models based on ALPs fulfill typically this bound, so that one 
needs not imposing it to guarantee compatibility with observations. 

%%%%%%%%%%%%%%%%%%%%%%%%%%%%%%%%%%%%%%%%%%%%%%%%%%%%%%%%%%%%%%%%%%%%%%%%%%%%%%%%%%%%%%%%%%%
\subsection{Axion-like particles}
\label{subsec:alps}

It is known that the inflaton field can be an ALP.
ALPs are hypothetical light particles, either scalar or pseudoscalar, associated with 
a shift symmetry and with derivative couplings to observed matter. These couplings 
are model dependent, but of particular 
importance is the interaction with a pair of photons, as given in~\cite{Arias:2012az},
\begin{align}
\label{eq:axion-photon_coupling}
\mathcal{L}_{\mathrm{int}}\supset-\frac{1}{4} g_{\phi\gamma\gamma} \, \phi 
\,F_{\mu\nu}\widetilde{F}^{\mu\nu},
\end{align}
where $ g_{\phi\gamma\gamma}$ is the coupling constant, $\phi$ is the ALP, 
$F^{\mu\nu}$ is the electromagnetic stress tensor and 
$\widetilde{F}^{\mu\nu}:= \epsilon^{\mu\nu\rho\sigma}F_{\rho\sigma}$ is 
its dual.
Unlike the QCD axion, 
ALPs do not necessarily solve the strong \CP problem
and, hence, they are subject to fewer constraints than the QCD 
axion~\cite{Ringwald:2014vqa,Masso:2006id,Masso:2002ip,Abbott:1982af}.

ALPs are natural to various extensions of the Standard Model (SM) of particle physics, 
such as those including the addition of extra \U1 or other symmetries. 
These extra symmetries are spontaneously broken by the nonzero vacuum expectation value (VEV) 
$\langle \psi \rangle=:v$ of some field $\psi$ charged under such symmetries. 
The ALP arises as the angular degree of freedom left after the breakdown of the 
symmetry. If the field $\psi(x)$ is parameterized as~\cite{Ringwald:2014vqa}
\begin{equation}
\label{eq:ALPparametrization}
\psi(x) ~=~ \left(v+\sigma(x)\right)e^{\I \phi(x)/f}\,,
\end{equation}
then $\phi$ is an ALP, $f$ its decay constant, and $\sigma$ is a small perturbation
close to the vacuum.

Due to the shift symmetry, evident in eq.~\eqref{eq:ALPparametrization}, the 
potential of an ALP is flat. However, if at a certain scale $\Lambda$ the interactions
related to the ALP become strong, some periodic potential is generated by e.g.\ instanton
effects. Such potential can take the form~\cite{Freese:1990rb}
\begin{equation}\label{eq:naturalinflationpotential}
    V ~=~ \Lambda ^4 \left[1 + \cos\left(\frac{\phi}{f}\right) \right] \,.
\end{equation}
Inflation driven by the potential~\eqref{eq:naturalinflationpotential} is known as natural 
inflation and has been widely studied before
(see e.g.~\cite{Freese:1990rb,Adams:1990pn,Adams:1992bn,
Yonekura:2014oja,Stein:2021uge}). 
To obtain a scalar spectral index value consistent with the observations,
one usually needs that\footnote{Super-Planckian decay constants are sometimes considered to be
a problem~\cite{Adams:1990pn, Banks:2003sx}, which we do not attempt to address here.} 
$f \gtrsim 5\, M_{\text{Pl}}$.

\begin{figure}[t]
        \includegraphics[width=\textwidth]{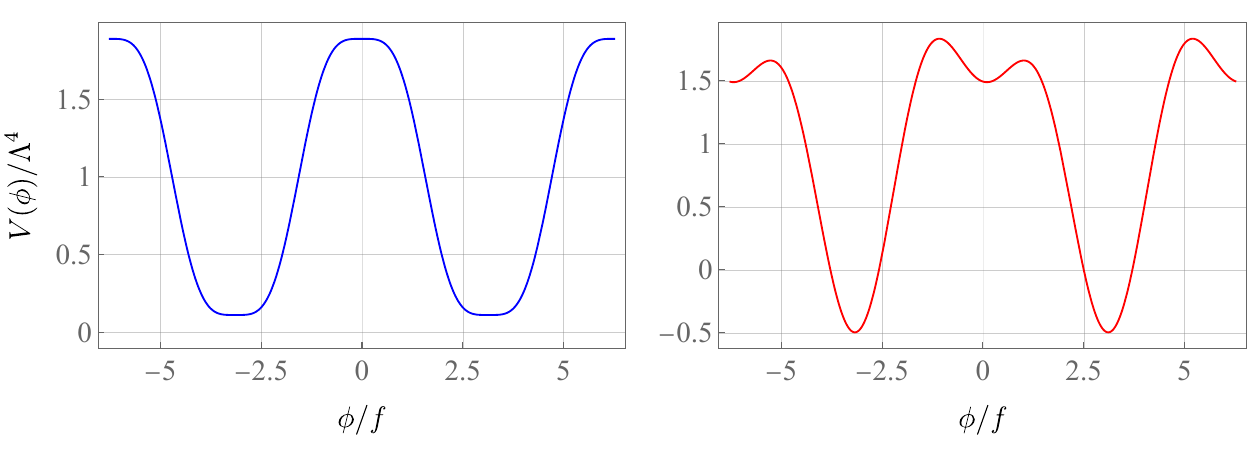}
 \caption{Hilltop potential~\eqref{eq:hilltop_potential} with different values of $n$,
    $\kappa$ and $\Theta$. For illustration, we choose $\kappa=1$, $n=3$, $\Theta=0$
    (left panel), and $\kappa=2$, $n=2$, $\Theta=0.1$ (right panel). When $\Theta=0$ 
    the potential displays the symmetry $V(\phi) = V(-\phi)$, 
    with $\kappa =1$ the maximum of the potential is flat, while the shape of the 
    minimum depends on $n$ and only two shapes are possible. \label{fig:hill-top}}
\end{figure}

Another example of inflation based on ALPs is hilltop inflation~\cite{Boubekeur:2005zm},
see e.g.~\cite{Daido:2017tbr,Daido:2017wwb} for more details.
The potential in this scenario has the form~\cite{Takahashi:2021tff}
\begin{equation}
\label{eq:hilltop_potential}
    V(\phi) ~=~ \Lambda^4\left[\cos \left(\frac{\phi}{f} + \Theta \right) -\frac{\kappa}{n^2}\cos\left(\frac{n\,\phi}{f}\right)\right] \, ,
\end{equation}
where the variation of the parameters $\kappa$, $\Theta$ and $n$ leads to 
changes in the shape of the potential. Some variations are shown in 
the figure~\ref{fig:hill-top}. The $\Lambda$ and $f$ parameters 
control the amplitude and the frequency of the potential, respectively.
The parameter $\kappa$ is related to the flatness at the top of the potential. 
When $\kappa=1$, the top of the potential is flat, and any other value yields a
local minimum at the top. Only when $\Theta = 0$, the potential exhibit the symmetry
$V(\phi) = V(-\phi)$.
When $n$ is odd, the field value at the minimum of the potential
closest to $\phi=0$, denoted by $\phi_{\text{min}}$, can be 
determined by the relation $\phi_{\text{min}}=\phi_{\text{max}} \pm \pi f$, where $\phi_{\text{max}}$
denotes the field at the maximum of the potential 
closest to $\phi=0$. In the left panel of the figure~\ref{fig:hill-top}, the values $\phi_{\text{max}}$, $\phi_{\text{min}}$
correspond to $\phi/f=0$ and $\phi/f=\pm \pi$, respectively. 
We consider $\phi_{\text{min}}>0$ without
loss of generality. Near zero and with $|\Theta| \ll 1 $, the potential
can be approximated as
\begin{align}
V(\phi)~\simeq~\frac{\Lambda^4}{f}\left[-\Theta \, \phi + \frac{\kappa - 1}{2 f}\,\phi^2
+\frac{\Theta}{6 f^2}\,\phi^3 - \frac{(\kappa\, n^2 -1)}{24 f^3}\,\phi^4\right] +\text{const.}
\end{align}
From the above expression, it is evident that when 
$\kappa \simeq 1$, the quadratic term is not relevant. Additionally,
the quartic term makes the largest contribution as $\phi \sim f$. For field values up to 
$\phi = \sqrt{6}\, f$, the linear term is dominant
over the cubic term. Therefore, the maximum of the potential
is approximated by solely considering the quartic and
linear terms. Using the relationship with $\phi_{\text{max}}$,
we have that
\begin{align}
\phi_{\text{min}}~\simeq~\pi f -\left[\frac{6 f^3 \,\Theta}{\kappa \, n^2-1}\right]^{\nicefrac13},
\end{align}
recalling that we assumed $\kappa \simeq 1$ and $n$ odd.
Nonzero values of $\Theta$ are known to produce spectral index values within
the observed range~\cite{Daido:2017wwb}. 

Arguably, a large decay constant might pose a problem due to the possible 
breakdown of the global axionic symmetry by quantum gravity effects~\cite{Kallosh:1995hi}.
In principle, this hurdle can also be addressed in axion-based inflationary
models~\cite{Kim:2004rp,Czerny:2014wza,Dimopoulos:2005ac}, and it has been even proposed 
that $f$ can be small in working examples~\cite{Daido:2017wwb,Daido:2017tbr}. However, as 
we show in appendix~\ref{sec:AppendixA}, avoiding large $f$ can lead to other more serious
phenomenological issues. Hence, we will assume here that the up-to-now unknown quantum 
gravitational effects do not destroy the structure of models based in axions.

\section{Building the model of dark inflaxion}
\label{sec:ALPXD}

To introduce our model, we start by discussing the features of the effective 
4-d action achieved when incorporating additional spatial dimensions. The scheme is 
inspired by the framework of dynamical dark matter and 
the interesting prospect of hilltop inflation~\cite{Daido:2017tbr} in top-down constructions. 

Theories that incorporate $D$ additional spatial dimensions result in an effective 4-d 
scheme after the process of compactification. Due to compactification, (4+$D$)-d fields 
decompose into an infinite set of 4-d states, which is commonly referred to as 
KK tower. The arrangement of compact dimensions determines the specific 
form of decomposition. Constraints on the compactification radius are established 
through different factors, including deviations in Newtonian gravity law, energy loss from supernovae, and 
the energy absorbed by a neutron-star~\cite{Workman:2022ynf}. 
Table~\ref{table:limitsRadius} displays the constraints on the upper limit of the compactification radius
for $D=1,2,3$ additional dimensions.

We consider one extra spatial dimension in which spacetime has the geometry of
$\mathcal{M}_4 \x S^1/\Z2$, and $x^M=(x^\mu, y)^\mathrm{T}$ denotes the coordinates. 
The coordinates $x^\mu$, $\mu=0,1,2,3$, describe the Minkowski spacetime $\mathcal{M}_4$, 
whereas $y\in\mathbbm{R}$ denotes the coordinate of the orbifolded extra dimension. 
The $S^1/\mathbb{Z}_2$ orbifold is obtained by moding out $y \sim y + 2\pi R$ 
to achieve an $S^1$ with radius $R$, and then the \Z2 isometry $y\sim -y$. 
These identifications imply that a 5-d field $\Phi=\Phi(x^\mu, y)$ must satisfy 
the boundary conditions $\Phi(x^\mu, y)=\Phi(x^\mu, y+2\pi \, R)$ and 
$\Phi(x^\mu,y)=\Phi(x^\mu,-y)$. Hence, in this case, a scalar field 
$\Phi$ possessing these properties is known to adopt the configuration of a
KK tower of the form~\cite{Dienes:1999gw}
\begin{align}
\label{eq:expansion}
\Phi(x^\mu,y) ~=~ \frac{1}{\sqrt{2\pi R}}\sum^{\infty}_{m=0}\,r_m 
\phi_m(x^\mu) \cos\left( \frac{m\, y}{R} \right)\,,
\end{align}
where $\phi_m(x^\mu)$ denote the 4-d states of the tower, and 
the coefficients are $r_0=1$ and $r_m=\sqrt{2}$
for $m>0$, in order to ensure the canonical normalization of the states.

\begin{table}[t]
\centering
\begin{tabular}{|c | c |  c |  c|} 
  \hline
 $D$ & 1 & 2 & 3 \\ 
 \hline\hline
 Torsion-balance & $2.22\times 10^{11}$ GeV$^{-1}$ & - & - \\   \hline
 SN 1987A & $2.48\times 10^{18}$ GeV$^{-1}$ & $4.86\times 10^9$ GeV$^{-1}$ & $5.77\times 
 10^6$ GeV$^{-1}$\\ \hline
 Neutron-star excess heat & $2.25\times 10^{11}$ GeV$^{-1}$ & $7.85\times 10^5$ GeV$^{-1}$& 
 $1.30\times 10^4$ GeV$^{-1}$\\ [.5ex] 
 \hline
\end{tabular}
\caption{Upper limits on the compactification radius of $D=1,2,3$ extra dimensions, determined
from torsion-balance experiments~\cite{Kapner:2006si}, 1987A supernova observations and 
neutron-star excess heat measurements~\cite{Hannestad:2003yd}.}
\label{table:limitsRadius}
\end{table}

On top of these considerations, the first crucial ingredient of our model 
is a massless 5-d axion-like field $\Phi$.
This field interacts with ordinary matter only via a derivative coupling, 
similar to the one displayed in eq.~\eqref{eq:axion-photon_coupling}. Furthermore, 
all 4-d SM fields are assumed to be confined to the subspace of the fixed point of the orbifold 
at $y=0$, and only the axion propagates throughout the full spacetime. 
Therefore, the axion and the SM particles only interact at $y=0$. 
Thus, the Lagrangian density for the 5-d field is given by
\begin{align}
\mathcal{L}_{5\text{-}\mathrm{d}} ~\supset~ 
\frac{1}{2}\partial^M \Phi \partial_M \Phi-
\delta(y)\frac{g_{\Phi\gamma\gamma}}{4}\,\Phi \,F_{\mu\nu}\widetilde{F}^{\mu\nu}\,,
\end{align}
where the coupling $g_{\Phi\gamma\gamma}$ has mass dimension $\nicefrac{-3}{2}$.
By expanding the 5-d field $\Phi$ in the tower of 4-d states, as in eq.~\eqref{eq:expansion} and integrating 
over the extra dimension, we arrive at the effective 4-d action
\begin{align}
\label{eq:4-d_action}
 S~\supset~\int d^4x\; \left(\frac{1}{2}\sum^{\infty}_{m=0}\partial^\mu 
 \phi_m \partial_\mu \phi_m -\frac{1}{2}\sum^{\infty}_{m=0}\frac{m^2}
 {R^2}\phi^2_m -\frac{g_{a\gamma\gamma}}{4}\sum^{\infty}_{m=0} r_m \phi_m 
 F_{\mu\nu}\widetilde{F}^{\mu\nu}\right)\,.
\end{align}
The 4-d coupling constant is defined as $g_{a\gamma\gamma}:=g_{\Phi\gamma\gamma}/\sqrt{2\pi R}$.

The second component of our model is a specific 5-d axion potential, that is assumed 
here to arise from non-perturbative effects with a natural scale $\Lambda$.
We consider the 5-d version of the hilltop potential in eq.~\eqref{eq:hilltop_potential}, 
which has the form
\begin{equation}
V_\mathrm{5-d} ~\supset~ \delta(y)\,\Lambda^4\left[\cos \left(\frac{\Phi}{f^{3/2}_{\Phi}} 
               + \Theta \right) -\frac{\kappa}{n^2}\cos\left(\frac{n\,\Phi}{f^{3/2}_{\Phi}}\right)\right]\,,
\end{equation}
where $f_{\Phi}$ has mass dimension $+1$. Using the decomposition of the 
field in its 4-d modes, the 4-d effective potential becomes
\begin{align}
\label{eq:4-d_potential}
V ~\supset~ \Lambda^4\left[ 
            \cos\left(\frac{\sum^{\infty}_{m=0}r_m  \phi_m}{f_a}+\Theta\right)
            -\frac{\kappa}{n^2}\cos\left(\frac{n\sum^{\infty}_{m=0}r_m  \phi_m}{f_a} \right) 
            \right]\,,
\end{align}
where the 4-d decay constant is $f_a:= \sqrt{2\pi R} f^{3/2}_\Phi$.
As the 4-d effective potential is built from the dynamics of a single 5-d field, 
$f_a$ is the sole decay constant and it is associated with all of the 4-d fields. 
Note that typical bottom-up models based on axions and with no extra dimensions 
naturally have multiple decay constants associated with different 4-d fields.

It can be assumed that there is a sector of KK modes that are so massive that 
they are decoupled from observable physics. Let us suppose that only $M+1$ fields
are undecoupled. After the heavier states decouple, we obtain the potential
of our {\it dark inflaxion} model:
\begin{align}
\label{eq:4-d_potentialN}
    V~=~\sum^{M}_{m=0}\frac{m^2}{2R^2}\phi^2_m+\Lambda^4\left[ 
\cos\left( \frac{\sum^{M}_{m=0}r_m\phi_m}{f_a} +\Theta\right) -\frac{\kappa}{n^2}
\cos\left( \frac{n\sum^{M}_{m=0}r_m\phi_m}{f_a} \right)\right].
\end{align}
Minimizing the potential leads to the field VEVs 
$\langle \phi_0 \rangle=(\pi \ell-[6\,\Theta/(\kappa\, n^2-1)]^{\nicefrac{1}{3}})f_a$ with $\ell \, \in 2\mathbb{Z}+1$,
and $\langle \phi_m \rangle=0$ for $m > 0$.
We study first the properties of the physical
4-d mass eigenstates. These states are constructed as in the framework of DDM.
The mass eigenstates, denoted by $\pmb{\psi}=(\psi_0, \psi_1, \ldots, \psi_M)^{\text{T}}$, 
are given by
\begin{align}
    \label{eq:Ppsi}
    \pmb{\psi}~=~P\,\pmb{\phi}.
\end{align}
Here, $P$ refers to the matrix that diagonalizes the squared mass matrix $\mathcal{M}^2$,
whose components are defined by
\begin{align}
    \mathcal{M}^2_{m m'} &:=\frac{\partial^2 V(\pmb{\phi})}{\partial \phi_m \partial \phi_{m'}}\nonumber\\
                         &=\frac{m^2}{R^2}\delta_{m m'}+
\frac{\Lambda^4}{f^2_a}r_mr_{m'}\left[ \kappa \cos\left( \frac{n \sum^{M}_{m''=0}r_{m''} \phi_{m''}}{f_a} \right) -\cos\left( \frac{\sum^{M}_{m''=0}r_{m''} \phi_{m''}}{f_a}+\Theta \right)\right].
\end{align}
Near the minimum of $V$, the squared mass matrix is
\begin{align}
\label{eq:massmatrix}
\mathcal{M}^2_{m m'}&~\simeq~\frac{m^2}{R^2}\delta_{m m'}+\frac{\Lambda^4}{f^2_a}r_m r_{m'}\left[ 1+\kappa \cos(n \pi) - \frac12\left(\frac{6 \Theta}{\kappa\, n^2-1}\right)^{\nicefrac23}\left(1+\kappa \, n^2 \cos (n\pi)\right)\right] \notag    \\
&~\simeq~M^2_c m^2\delta_{m m'}+m^2_\Lambda\,r_m r_{m'} \,.
\end{align}
In the last equation, we have set $n=3$ and $\kappa = 1$. The 
compactification scale $M_c$ and the natural scale of the axion $m_\Lambda$ are 
introduced and defined here as
\begin{align}
\label{eq:m_lambda}
    M^2_c         ~:=~ \frac{1}{R^2} \qquad \text{and}\qquad
    m^2_{\Lambda} ~:=~ (6\Theta)^{\nicefrac23}\frac{\Lambda^4}{f^2_a}\,.
\end{align}
As we will discuss in section~\ref{sec:benchmark}, the physics of inflation
and DM are linked to relations between these two scales.

As shown in~\cite{Daido:2017wwb}, the hilltop potential presented in 
eq.~\eqref{eq:hilltop_potential} has the capability to yield suitable values 
of spectral index and tensor-to-scalar ratio during inflation.
Accordingly, we aim at formulating the second term of the potential in 
eq.~\eqref{eq:4-d_potentialN}, related to the hilltop-inflation model, as a 
function of a single axion-like field. This field could act as an inflaton, 
provided that the scale of the hilltop term is larger than the scale of the 
quadratic term. To accomplish this configuration, we perform a rotation in 
field space, assuming that precisely the linear combination of the $M+1$ 
$\phi_m$ fields in the cosine is an effective mass eigenstate.

A general rotation in the field space is given by
\begin{equation}
\label{eq:change_basis}
\pmb{\phi}=\mathcal{R}(\pmb{\varphi})\pmb{a},
\end{equation}
where $\pmb{\phi}=(\phi_0,\phi_1,\dots, \phi_M)^T$, $\mathcal{R}
(\pmb{\varphi})$ is an element 
of the group of rotations in $M+1$ dimensions, and $\pmb{a}=
(a_0,a_1,\dots,a_M)^T$ is the new basis. 
We denote the mass eigenstates as $\pmb{\psi}$, resulting from the 
diagonalization of the square mass matrix $\mathcal{M}^2$, and the 
rotated states as $\pmb{a}$. As shown later, there exists a particular
scenario of the ratio between the $M_c$ and $m_\Lambda$ scales in which
the $\pmb{a}$ states and the $\pmb{\psi}$ are equivalent.
The conditions for having the hilltop term as a function of a single field, 
say $a_M$, are 
\begin{align}
\label{eq:restrictions}
    \sum^{M}_{m=0}r_m \mathcal{R}_{mj}~\stackrel!=~0, \qquad \forall \, j \neq M\,,
\end{align}
so that the coefficients of the fields $a_0, a_1, \ldots, a_{M-1}$ 
vanish in the cosine. This condition implies that
\begin{align}
\label{eq:restrictions2}
\sum^{M}_{m=0}r_m\phi_m~=~ a_M\sum^{M}_{m=0}r_m\mathcal{R}_{mM}\,,
\end{align}
where there is no sum over the index $M$ on the right-hand side. This condition 
implies that the direction $a_M$ of the basis $\pmb{a}$ can become the 
hilltop-like direction of the potential.
The potential in eq.~\eqref{eq:4-d_potentialN}, after imposing the
conditions in eq.~\eqref{eq:restrictions}, has the form
\begin{align}
    \notag V~=~&\sum^{M}_{m=0}\frac{m^2}{2R^2}\left( \sum^M_{j=0} \mathcal{R}_{mj}a_j \right)^2\\
    &+\Lambda^4\left[ \cos\left( \frac{a_M\sum^{M}_{m=0}r_m\mathcal{R}_{mM}}{f_a} +\Theta\right) -\frac{\kappa}
{n^2}\cos\left( \frac{n\,a_M\sum^{M}_{m=0}r_m\mathcal{R}_{mM}}{f_a} \right)\right].
\end{align}
The couplings of the new states with a pair of photons then become
\begin{align}
\label{eq:int_rot}
\mathcal{L}_{\mathrm{int}}=-\frac{g_{a\gamma\gamma}}{4}a_M F_{\mu\nu} \widetilde{F}^{\mu\nu}\sum^{M}_{m=0}r_m \mathcal{R}_{mM}.
\end{align}
The restrictions of the rotation matrix eq.~\eqref{eq:restrictions} indicate 
that only $a_M$ interacts with photons. Specific cases associated with this promising 
basis choice will be analyzed. We shall show that this choice is equivalent to
considering $\Lambda\gg R$ in the actual mass eigenstates.

Using these features, we will now study different realizations of dark inflaxion 
depending on the quantity of low-energy fields available in our model. The dynamics 
of the inflaton and DM will be set by a series of correlations between $M_c$
and $m_\Lambda$.

\section{Benchmark scenarios}
\label{sec:benchmark}

Let us now study 
three sample cases, in which one, two and three fields are physically relevant
and all others are decoupled. We shall perform an appropriate basis transformation,
which can conveniently lead to one promising scenario, where one single 4-d 
axion-like field couples to photons.

\subsection{Single-field}
\label{subsec:1field}

In the extreme case of a single field, $M=0$, all massive states of the KK tower
are decoupled. Consequently, the potential is expressed as in
eq.~\eqref{eq:hilltop_potential}, with the identifications 
$\phi_0 \rightarrow \phi$ and $f_a \rightarrow f$. As there is no 
mass term in the KK tower in 4-d action~\eqref{eq:4-d_action} for the zero mode, 
this situation leads to the analysis carried out in the classical model of
axion hilltop inflation~\cite{Daido:2017wwb}.

%%%%%%%%%%%%%%%%%%%%%%%%%%%%%%%%%%%%%%%%%%%%%%%%%%%%%%%%%%%%%%%%%%%%%%%%%%%%%%%%%%%%%%%%%%%%%%%%%%%%%%%%%%%%%%%%
\subsection{Two fields}
\label{subsec:2fields}

Scenarios with $M>1$ fields are considered a generalization of the 
two-field case.
\begin{figure}[t]
   \includegraphics[scale=.5]{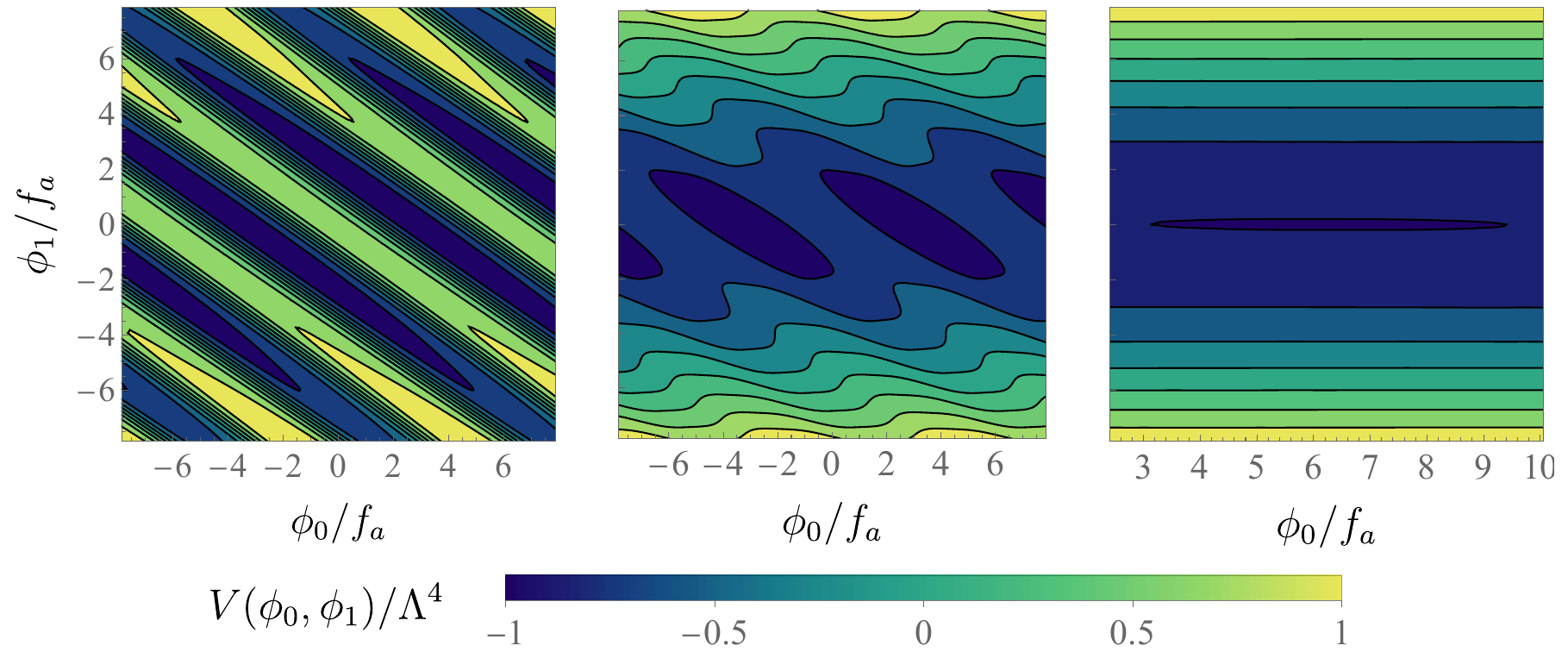}
   \caption{Potential~\eqref{eq:unrotated_potential} with $\kappa=1$,
    $n=3$, $\Theta=0.015$, $f_a=2 \Lambda$ and different hierarchies 
    between $M_c$ and $m_\Lambda$. In the left panel $R=10^{3}\Lambda^{-1}$,
    yielding $m^2_\Lambda\gg M^2_c$; in the center panel $R={2^{1/6}f_a}/{3^{\nicefrac13}\Theta^{1/6}\Lambda^2}$ 
    corresponding to $m^2_\Lambda=M^2_c$; and for the right panel 
    $R=10^{-3}\Lambda^{-1}$ associated with $m^2_\Lambda \ll M^2_c$.
    \label{fig:two_field_potential}}
\end{figure}
The two-field potential has the structure
\begin{align}
\label{eq:unrotated_potential}
  V~=~\frac{1}{2R^2}\phi^2_1+\Lambda^4\left[ \cos\left( 
\frac{\phi_0+\sqrt{2}\phi_1}{f_a} +\Theta\right) -\frac{\kappa}{n^2}\cos
\left( \frac{n(\phi_0+\sqrt{2}\phi_1)}{f_a} \right)\right],
\end{align}
which is shown in figure~\ref{fig:two_field_potential}.
We plot the potential for $\kappa = 1$, $n=3$, $f_a=2\Lambda$ and $\Theta=0.015$, and 
various ratios of the axion natural scale $m_\Lambda$ to the compactification scale $M_c$. 
A general rotation in the two-field space takes the form
\begin{align}
\label{eq:2dBasisRotation}
\begin{pmatrix}
\phi_0 \\
\phi_1
\end{pmatrix}=~
\mathcal{R}(\varphi)\begin{pmatrix}
a_0 \\
a_1
\end{pmatrix}
=
\begin{pmatrix}
\cos \varphi  & \sin \varphi\\
-\sin \varphi & \cos \varphi
\end{pmatrix}
\begin{pmatrix}
a_0 \\
a_1
\end{pmatrix},
\end{align}
where $\varphi \in [-\pi, \pi]$. 
The conditions stated in eq.~\eqref{eq:restrictions} imply $\cos \varphi- \sqrt{2} \sin \varphi = 0$, 
thus resulting in
\begin{align}
\label{eq:rot_fields}
    \begin{pmatrix}
        a_0 \\
        a_1
    \end{pmatrix}
    ~=~
    \frac{1}{\sqrt{3}}
    \begin{pmatrix}
        \sqrt{2}\phi_0  -\phi_1\\
        \phi_0+ \sqrt{2}\phi_1
    \end{pmatrix}\,.
\end{align}
The potential in the rotated basis is hence given by
\begin{align}
\label{eq:rotated_potential}
V~=~\frac{1}{6R^2}\left(\sqrt{2}a_1-a_0 \right)^2
      +\Lambda^4\left[ \cos\left(\frac{\sqrt{3}\,a_1}{f_a} +\Theta\right) 
      -\frac{\kappa}{n^2}\cos\left( \frac{n\sqrt{3}\,a_1}{f_a} \right)\right].
\end{align}
The inflaton is the only field that interacts with photons in the new basis, as 
specified by eq.~\eqref{eq:int_rot}. This interaction in this case is
\begin{align}
\label{eq:coupling}
\mathcal{L}_\mathrm{int} ~=~ -\frac{\sqrt{3}}{4}g_{a\gamma\gamma} a_1F_{\mu\nu}\widetilde{F}^{\mu\nu}\,.
\end{align}
Therefore, the full effective axion-like action in the rotated basis becomes
\begin{align}
\label{eq:actionRotatedCase}
\notag  S~\supset~\int d^4x\; &\left\{\frac{1}{2}\sum^{1}_{i=0}\partial^\mu 
 a_i \partial_\mu a_i -\frac{1}{6R^2}\left( \sqrt{2}a_1-a_0 \right)^2\right.\\
&\left.-\Lambda^4\left[ \cos\left(\frac{\sqrt{3}\,a_1}{f_a} + \Theta\right) 
       -\frac{\kappa}{n^2}\cos\left( \frac{n\sqrt{3}\,a_1}{f_a} \right)\right]
       -\frac{\sqrt{3}}{4}g_{a\gamma\gamma} a_1F_{\mu\nu}\widetilde{F}^{\mu\nu}\right\}\,.
\end{align}

Due to its hilltop couplings, $a_1$ is assumed to play the role of the inflaton.
This is further enabled by setting e.g.\ $R=10^3\Lambda^{-1}$, 
which for a nonzero value of $\Theta$ ensures that $m_\Lambda\gg M_c$.
The mass of the inflaton at the potential minimum is given by
${m}^2_{a_1}:=V''({a_1}_\mathrm{min}) =\partial^2V({a_1}_\mathrm{min})/\partial {a_1}^2$,
where ${a_1}_\mathrm{min}$ denotes the value of the field at the minimum of 
the potential.
The hilltop potential has an upside-down symmetry when the value of
$n$ is odd, resulting in $V''({a_1}_{\min})=-V''({a_1}_{\max})$, as stated 
in section~\ref{sec:inflationalps}. 
In this way, we can use the maximum value of the field and evaluate it in 
the curvature of the potential to obtain the mass of $a_1$.

Using a similar procedure to that outlined in section~\ref{subsec:alps}, 
the evaluation of the $a_1$ field at the maximum potential leads to 
\begin{align}
{a_1}_{\max}~\simeq~-\left[\frac{2 f^3_a \,\Theta}{\sqrt{3}(\kappa\, n^2-1)}\right]^{\nicefrac13}.
\end{align}
The squared mass of the field $a_0$ is given by $m^2_{a_0}:=\partial^2V/\partial {a_0}^2$. 
By using the value of ${a_1}_{\text{max}}$ to determine $V''({a_1}_{\text{min}})$,
the masses of the fields are
\begin{subequations}
\label{eq:mass}
\begin{align}
\label{eq:ma0}
m_{a_0}^2 ~=&~\,\frac{1}{3R^2}\,,\\
\label{eq: ma1}
m_{a_1}^2~\simeq&~\frac{2}{3R^2} 
                + 3\frac{\Lambda^4}{f^2_a}\left[1-\kappa+\frac12(6\, \Theta)^{\nicefrac23}(\kappa\, n^2-1)^{\nicefrac13}\right] \notag \\
             ~= &~\frac{2}{3R^2} + 3(6\Theta)^{\nicefrac23}\frac{\Lambda^4}{f^2_a} \,,
\end{align}
\end{subequations}
where in the last equation we set $\kappa = 1$ and $n=3$.

%%%%%%%%%%%%%%%%%%%%%%%%%%%%%%%%%%%%%%%%%%%%%%%%%%%%%%

\subsubsection{General mass eigenstates}

The discussed rotated states represent a particular choice of mass eigenstates, 
where the fundamental scale of the axion is significantly larger than the compactification 
scale. We will therefore proceed with the construction of general mass eigenstates under the 
assumption that $n=3$ and $\kappa=1$. Following eq.~\eqref{eq:Ppsi}, the mass eigenstates are 
constructed from the matrix $P$ that diagonalizes the squared mass matrix $\mathcal{M}^2$. 
With the components of $\mathcal{M}^2$ given by eq~\eqref{eq:massmatrix}, in the two-field 
case, we find
\begin{align}
    P~=~
    \begin{pmatrix}
        \frac{-c_+}{\sqrt{8m^4_\Lambda+c^2_+}} & \frac{\sqrt{8}m^2_\Lambda}{\sqrt{8m^4_\Lambda+c^2_+}}\\
        \frac{-c_-}{\sqrt{8m^4_\Lambda+c^2_-}} & \frac{\sqrt{8}m^2_\Lambda}{\sqrt{8m^4_\Lambda+c^2_-}}
    \end{pmatrix},
    \end{align}
where
\begin{align}
\label{eq:cpm}
    c_{\pm}~:=~M^2_c+m^2_\Lambda \pm \sqrt{M^4_c+2M^2_cm^2_\Lambda+9m^4_\Lambda}\,.
\end{align}
Hence, the mass eigenstates are
\begin{align}
\label{eq:eigenstates}
\begin{pmatrix}
    \psi_0 \\
    \psi_1\\
    \end{pmatrix}
    ~=~
    \begin{pmatrix}
        \frac{1}{\sqrt{8m^4_\Lambda+c^2_+}}\left[-c_+\phi_0 + \sqrt{8}m^2_\Lambda\phi_1\right]\\
        \frac{1}{\sqrt{8m^4_\Lambda+c^2_-}}\left[-c_-\phi_0 + \sqrt{8}m^2_\Lambda\phi_1\right]\\        
        \end{pmatrix},
\end{align}
whose interactions with photons are given by
\begin{align}
\label{eq:interactions}
    \notag\mathcal{L}_\mathrm{int}~=~&\left[ \frac{-\sqrt{8m^4_\Lambda+c^2_+}}{\sqrt{2}}-\frac{c_{-}}{2\sqrt{2}}\left( \sqrt{\nicefrac{M^4_c}{m^4_\Lambda}+2\nicefrac{M^2_c}{m^2_\Lambda}+9} +\frac{M^2_c}{m^2_\Lambda}+1\right) \right]\psi_0 F_{\mu\nu}\widetilde{F}^{\mu\nu}\\
    +&\left[ \frac{\sqrt{8m^4_\Lambda+c^2_{-}}}{\sqrt{2}}+\frac{c_{+}}{2\sqrt{2}}\left( \sqrt{\nicefrac{M^4_c}{m^4_\Lambda}+2\nicefrac{M^2_c}{m^2_\Lambda}+9} -\frac{M^2_c}{m^2_\Lambda}-1\right) \right]\psi_1 F_{\mu\nu}\widetilde{F}^{\mu\nu}\,.
\end{align}
In our general analysis, the heaviest field is denoted as $\psi_1$.

\begin{figure}[t]
\centering
   \includegraphics[scale=.75]{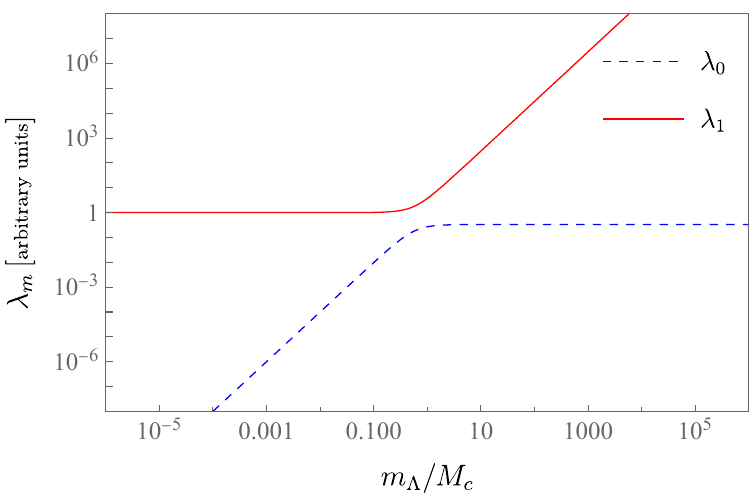}
   \caption{Squared-mass eigenvalues, $\lambda_0$ and $\lambda_1$, of $\mathcal{M}^2$ eigenstates
            $\psi_0$ and $\psi_1$.
            In the scenario where $m_\Lambda \ll M_c$, $\lambda_0$ is exceptionally small compared 
            with $\lambda_1$. When $m_\Lambda\sim M_c$, both masses are similar. If $m_\Lambda\gg M_c$, 
            then $\lambda_1$ is much larger than $\lambda_0$.
            \label{fig:twoeigenvalues}}
\end{figure}

From the numerical values $c_{\pm}$ in eq.~\eqref{eq:cpm}, we observe that the mass eigenstates 
exhibit distinct orientations in field space depending on the ratio ${m_\Lambda}/{M_c}$. 
These orientations appear in three configurations based on whether the natural axion 
scale is much larger, equal or much smaller than the compactification scale. 
Figure~\ref{fig:twoeigenvalues} illustrates the squared masses of the $\psi_0$ and $\psi_1$ 
fields, corresponding to the eigenvalues $\lambda_0$ and $\lambda_1$ of the squared mass matrix 
$\mathcal{M}^2$. As we shall shortly discuss in more detail, when ${m_\Lambda}/{M_c}\ll 1$, the mass of the $\psi_1$ field is of the 
order of $M_c$, whereas $\psi_0$ is significantly lighter. If ${m_\Lambda}/{M_c}\sim1$, then both masses 
are comparable. For ${m_\Lambda}/{M_c} \gg 1$, the mass of $\psi_0$ is 
proportional to $M_c$, whereas $\psi_1$ is significantly heavier.

Given the variation resulting from the ratio of the scales, we investigate the structure of the 
mass eigenstates and their interactions, shown in eqs.~\eqref{eq:eigenstates} and~\eqref{eq:interactions},
respectively, for ${m_\Lambda}/{M_c}$ significantly higher, equivalent, or considerably lower than one.

\subsubsection[m\_Lambda >> Mc]{\boldmath$m_{\Lambda} \gg M_c$\unboldmath}
\label{subsubsec:ddi}

When the natural scale of the axion is significantly greater than the compactification scale, 
the mass eigenstates are
\begin{align}
    \begin{pmatrix}
        \psi_0\\
        \psi_1\\
    \end{pmatrix}~=~
    \frac{1}{\sqrt{3}}
        \begin{pmatrix}
        -\sqrt{2}\phi_0 +(1-\nicefrac{M^2_c}{3m^2_\Lambda})\phi_1 \\
        \phi_0+\sqrt{2}(1+\nicefrac{M^2_c}{3m^2_\Lambda})\phi_1\\
    \end{pmatrix}
    ~\simeq~
    \frac{1}{\sqrt{3}}\begin{pmatrix}
        -\sqrt{2}\phi_0 + \phi_1\\
        \phi_0+\sqrt{2}\phi_1\\
    \end{pmatrix},
\end{align}
and the squared masses of the fields are the corresponding eigenvalues,
\begin{align}
    m^2_{\psi_0}~\approx~\frac{1}{3}M^2_c \qquad\text{and} \qquad 
    m^2_{\psi_1}~\approx~3m^2_\Lambda\,.
\end{align}
The interaction Lagrangian as a function of mass eigenstates becomes
\begin{align}
\label{eq:int1}
    \notag \mathcal{L}_\mathrm{int}&~=~ 
      \frac{\sqrt{3}}{2\sqrt{2}}\left( \frac{M_c}{3m_\Lambda}\right)^2 g_{a\gamma\gamma}\psi_0 F_{\mu\nu}\widetilde{F}^{\mu\nu}
     +\frac{\sqrt{3}}{4}\left[\left( \frac{M_c}{3m_\Lambda}\right)^2 -1\right]g_{a\gamma\gamma}\psi_1 F_{\mu\nu}\widetilde{F}^{\mu\nu}\\
     &~\simeq~-\frac{\sqrt{3}}{4}g_{a\gamma\gamma}\psi_1 F_{\mu\nu}\widetilde{F}^{\mu\nu}\,.
\end{align}

We observe that in the vanishing limit of this scenario, i.e.\ as 
${M_c}/{m_\Lambda} \rightarrow 0$, 
the mass eigenstates and the rotated states in eq.~\eqref{eq:rot_fields}
prove to be equivalent. Specifically, we see that we can establish the 
identifications $a_0 = -\psi_0$ and $a_1 = \psi_1$ by noting the masses
and interactions. 
However, as shown in eq.~\eqref{eq:int1}, there remains a coupling between the 
$\psi_0$ field and a photon pair in the mass eigenstates. Despite being 
suppressed by the factor ${M^2_c}/{m^2_\Lambda}$, 
this coupling may have some physical relevance.

\subsubsection[m\_Lambda = Mc]{\boldmath $m_{\Lambda} = M_c$ \unboldmath}

For the scenario where  $M_c=m_\Lambda$, the axion-like mass eigenstates are
\begin{align}
    \begin{pmatrix}
        \psi_0\\
        \psi_1\\
    \end{pmatrix}~=~
    \frac{\sqrt{3+\sqrt{3}}}{\sqrt{6}}
        \begin{pmatrix}
        -\phi_0 + \left(\sqrt{2-\sqrt{3}}\right)\phi_1 \\
       \left(\sqrt{2-\sqrt{3}}\right) \phi_0+\phi_1\\
    \end{pmatrix},
\end{align}
and the corresponding eigenvalues are
\begin{align}
    m^2_{\psi_0}~=~(2-\sqrt{3})m^2_\Lambda 
    \qquad\text{and}\qquad 
    m^2_{\psi_1}~=~(2+\sqrt{3})m^2_\Lambda\,.
\end{align}
The axion-photon interactions are governed by
\begin{align}
    \mathcal{L}_\mathrm{int}~=~ \frac{\left( 3-\sqrt{3} \right)^{3/2}}{24}g_{a\gamma\gamma}\psi_0
 F_{\mu\nu}\widetilde{F}^{\mu\nu}-\frac{\left( 3+\sqrt{3} \right)^{3/2}}{24}g_{a\gamma\gamma}\psi_1
 F_{\mu\nu}\widetilde{F}^{\mu\nu}\,.
\end{align}
Note that the interaction between the $\psi_0$ field and a pair of photons 
remains present and is not suppressed, unlike the previous case.

\subsubsection[m\_Lambda << Mc]{\boldmath$m_{\Lambda} \ll M_c$ \unboldmath}
\label{subsubsec:BigMc}

The eigenstates of our two-field dark inflaxion model are given by
\begin{align}
    \begin{pmatrix}
        \psi_0\\
        \psi_1\\
    \end{pmatrix}~=~
        \begin{pmatrix}
        -\phi_0 +\frac{m^2_\Lambda}{M^2_c}\phi_1 \\
        \frac{3}{2}\frac{m^2_\Lambda}{M^2_c}\phi_0+\left( 1+\frac{m^2_\Lambda}{M^2_c} \right)\phi_1\\
    \end{pmatrix}~\simeq~
    \begin{pmatrix}
        -\phi_0 \\
            \phi_1
    \end{pmatrix}\,,
\end{align}
and their eigenvalues are
\begin{align}
    m^2_{\psi_0}~\approx~m^2_\Lambda \qquad\text{and}\qquad 
    m^2_{\psi_1}~\approx~M^2_c\,.
\end{align}
Further, the associated axion-photon interaction Lagrangian is
\begin{align}
    \mathcal{L}_\mathrm{int} ~\simeq~ \frac{1}{4}g_{a\gamma\gamma}\psi_0 F_{\mu\nu}\widetilde{F}^{\mu\nu}-
\frac{\sqrt{2}}{4}g_{a\gamma\gamma}\psi_1 F_{\mu\nu}\widetilde{F}^{\mu\nu}\,.
\end{align}

\subsubsection{Some cosmological constraints}
\label{sec:CosmologicalExclusion}

\begin{figure}[b!]
\centering
   \includegraphics[scale=.7]{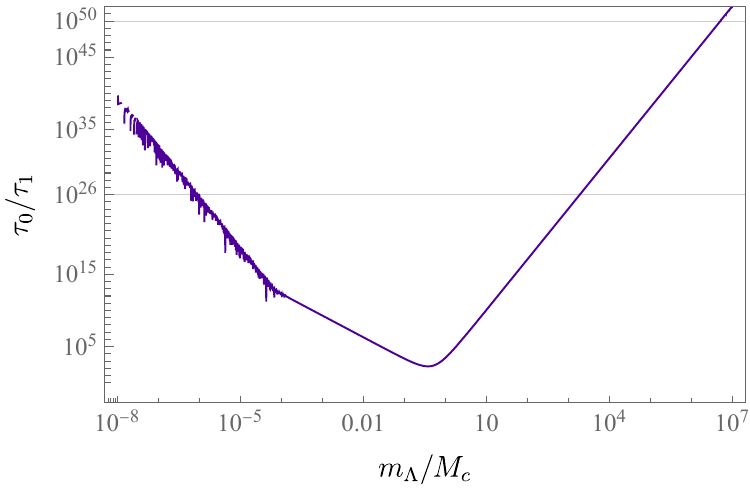}
   \caption{Ratio of the mean lifetimes of $\psi_0$ and $\psi_1$
           in terms of the ratio $m_\Lambda$ to $M_c$. The two horizontal lines,
           associated with $\tau_0/\tau_1 = 10^{26}$ and  $\tau_0/\tau_1 = 10^{50}$,
           correspond to the minimum reheating temperature considered. The lower line 
           is derived from considering $T_\mathrm{RH}\gtrsim 1$\,MeV to produce sufficient 
           reheating. The upper line comes from considering $T_\mathrm{RH}\gtrsim 10^{9}$\,GeV, 
           using the baryon asymmetry as a restriction for the reheating temperature~\cite{Kolb:1990vq}.
           \label{fig:ratio}}
\end{figure}

In general, the three limiting cases can be further cosmologically constrained if one insists 
on demanding that one the fields be the inflaton while the other play the role of a
DM candidate.

First, if the lighter field $\psi_0$ constitutes all of the DM, it is necessary for its mean 
lifetime to fulfill $\tau_0\gtrsim 10^{26}$\,s~\cite{Zhang:2007zzh}.

Secondly, there are constraints on the mean lifetime $\tau_1$ of the field $\psi_1$ that 
is considered to be an inflaton, see ref.~\cite{Kolb:1990vq}. The {\it sufficient reheating condition} 
requires the Universe to be radiation-dominated during Big Bang Nucleosyntesis. Thus, 
it is necessary for the reheating temperature to be $T_\mathrm{RH} \gtrsim 1$\,MeV. 
Additionally, the relation between this temperature and the inflaton mean lifetime,
$T_\mathrm{RH}^2=\tau^{-1}_1 M_\mathrm{Pl}$, implies that 
$\tau_1\lesssim 1$\,s. Furthermore, the limiting factor on $T_\mathrm{RH}$
due to the baryon asymmetry requires it to be larger than $10^{9}$\,GeV, leading to 
$\tau_1\lesssim 10^{-24}$\,s. 

From these considerations, we conclude that the ratio of the mean lifetimes of the fields must 
at least comply with $\tau_0/\tau_1 \gtrsim 10^{26}$ for enough reheating to happen. 
If one further relates the baryon asymmetry to the reheating temperature, one might even demand 
$\tau_0/\tau_1 \gtrsim 10^{50}$. The figure~\ref{fig:ratio} illustrates
the ratio of mean lifetimes as a function of the ratio between $m_\Lambda$ and $M_c$ scales. 
For a two-field model, the blue curve of $\tau_0/\tau_1$ must be over the bounding 
horizontal lines. The figure reveals that the possibility of dark inflaxion with 
$m_\Lambda \sim M_c$, i.e.\ inflation and DM arising simultaneously from a single 5-d
ALP with degenerate masses, is excluded. Notice though that this case could yield
other phenomenologically viable scenarios, such as DDM.

This restricts the dark inflaxion scenarios to those in which $m_\Lambda \gg M_c$ and $m_\Lambda \ll M_c$.
In the latter case, the original KK states $\phi_0$ and $\phi_1$ are equivalent to the mass eigenstates
$\psi_0$ and $\psi_1$. This implies that the potential of the mass eigenstates takes its full
shape~\eqref{eq:unrotated_potential}, where the combination of states in the hilltop terms
suggests a more complex multifield inflationary model. The intricate dynamics of this phenomenon 
will be examined elsewhere.
On the other hand, we notice that figure~\ref{fig:ratio} seems to restrict somewhat less 
the case $m_\Lambda \gg M_c$, which, as mentioned earlier, is equivalent to the convenient 
basis choice leading to the simple potential~\eqref{eq:rotated_potential} for a single 4-d
axion-like inflaton. For this reason, we shall focus on this more promising and simpler 
scenario.

\subsubsection{Initial conditions}

We will consider the initial conditions as in the dynamic dark matter scenario~\cite{Dienes:2011ja}. 
We assume that, prior to the emergence of non-perturbative effects that trigger its
dynamics, the 5-d field $\Phi$ is massless and it is assigned a random initial value. 
This value can be parameterized as
\begin{equation}
\langle \Phi \rangle ~=~ \theta f^{3/2}_\Phi\,,
\end{equation}
where $\theta \in [-\pi,\pi]$.
After dimensional reduction, the only 4-d field with a non-zero expectation value is 
the zero mode, which holds the value
\begin{align}
\langle \phi_0 \rangle ~=~ \sqrt{2\pi R} \,\theta f^{3/2}_\Phi\,.
\end{align}
After the change of basis~\eqref{eq:2dBasisRotation}, the expected 
values of the rotated fields are
\begin{align}
\langle a_m \rangle ~=~\sum^1_{n=0}\langle \phi_n \rangle \langle \phi_n | a_m \rangle\,, \qquad m=0,1.
                    \end{align}
where the projection $\langle \phi_n | a_m \rangle$ is rotation matrix element $R_{nm}$. As the expected value of 
$\phi_n$ fields is only non-zero for $\langle \phi_0 \rangle$, it follows that 
\begin{align}
\langle a_m \rangle ~=~\langle \phi_0 \rangle R_{0m}, \qquad m=0,1.
\end{align}
The matrix elements are
\begin{align}
\label{eq:initial_conditions}
R_{00}~=~\sqrt{2}R_{01}~=~\sqrt{\frac{2}{3}}\,,
\end{align}
which means that the initial values are related by
\begin{align}
\label{eq:initial_values}
\langle a_0 \rangle ~=~ \sqrt{2}\,\langle a_1 \rangle\,.
\end{align}

%%%%%%%%%%%%%%%%%%%%%%%%%%%%%%%%%%%%%%%%%%%%%%%%%%%%%%%%%%%%%%%%%%%%%%%%%%%%%%%%%%%%%%%%%%%%%%%%%%%%%%%%%%%%%%%%
\subsection{Multiple fields}
\label{subsec:Mfields}

\begin{figure}[b!]
\centering
   \includegraphics[scale=.75]{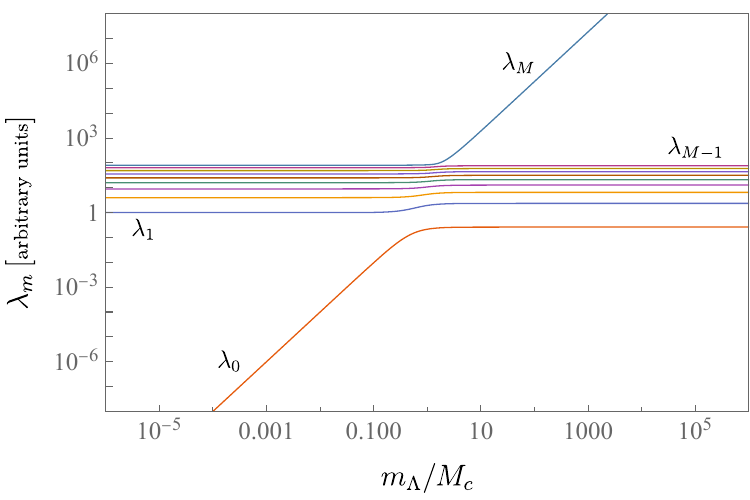}
   \caption{Squared-mass eigenvalues, $\lambda_n$, of $\mathcal{M}^2$ eigenstates $\psi_n$.
            In the scenario where $m_\Lambda \ll M_c$, $\psi_0$ is exceptionally light, 
            while the remaining fields have quasi-degenerate and significantly larger masses.
            When $m_\Lambda\sim M_c$, all masses are similar. If $m_\Lambda\gg M_c$, 
            then $\psi_M$ is very heavy, and the remaining masses are smaller and quasi-degenerate.
            \label{fig:Meigenvalues}}
\end{figure}

For illustrative purposes, to establish what kind of mass structure we should expect 
in the more general case of undecoupled 4-d matter, 
we have explored how the mass eigenstates are distributed if one considers the extended 
case with $M+1$ undecoupled axion-like fields.
Figure~\ref{fig:Meigenvalues} displays the eigenvalues associated with $\psi_m$ mass eigenstates,
denoted by $\lambda_m$, corresponding to their squared masses. As in the simplest two-field
framework, we observe three limiting scenarios: where $m_\Lambda/M_c$ is significantly smaller, 
equal, or significantly larger than one. 

For $m_\Lambda/M_c\ll1$, we note that one eigenstate is very 
light while all the others have similar masses of order $M_c$, differing by up to one 
order of magnitude. The case $m_\Lambda/M_c\sim1$ yields, as expected,
the masses of all $M+1$ fields are comparable, and they exhibit a 
scaling relationship of $\lambda_m \propto (m+1)^2 M^2_c$, with $0 \leq m \leq  M$.
All of the fields couple to photons, so that their dynamics is very similar.
As explained earlier in section~\ref{sec:CosmologicalExclusion}, this case do not
yield dark inflaxion. However, we expect this scenario to yield either a multifield 
inflationary model or a DDM-like case with a yet unexplored potential.
Finally, if $m_\Lambda/M_c\gg1$, there are $M$ fields with comparable masses
of order $M_c$ and one extremely heavy field. One arrives easily to the simplest 
dark inflaxion case, where the inflaton is $\psi_M$ and the remaining fields build 
a multifield DM sector.

This discussion shows that our two-field discussion straightforwardly generalizes to
the most general case and is, hence, sufficient to illustrate the phenomenological
potential of our model.

\section{Dynamics of the inflaton and DM in dark inflaxion}
\label{sec:dynamics}

We are finally ready to analyze the details of the simplest case of dark inflaxion, 
in which the natural scale of the axion is much larger than the compactification scale, 
$m_\Lambda/M_c \gg 1$. Recall that, as shown in section~\eqref{subsubsec:ddi}, the 
rotated states $a_0$ and $a_1$, are equivalent to the mass eigenstates $\psi_0$ and 
$\psi_1$ in this case. So, we can use the effective action in eq.~\eqref{eq:actionRotatedCase}.
In this case, only the inflaton $a_1$ interacts with photons while $a_0$ remains 
stable. Furthermore, due to the hierarchy of their masses, their dynamics are decoupled.

Considering the fields $a_m$ to be spatially homogeneous, the equations of motion are 
\begin{align}
\label{eq:motion_equation}
\ddot{a}_m+\left(3H(t)+\Gamma_m\right)\dot{a}_m+\frac{\partial V(\pmb{a})}{\partial a_m}~=~0\,, 
\qquad m=0, 1,
\end{align}
where, as a consequence of eq.~\eqref{eq:coupling}, we have\footnote{Note that, in
general, $\Gamma_0\neq0$ is suppressed by $(M_c/m_\Lambda)^2$, see eq.~\eqref{eq:int1}. 
DM stability however can be controlled by the size of $g_{a\gamma\gamma}$ and the fact 
that $(M_c/m_\Lambda)^2\lesssim10^{-6}$, see figure~\ref{fig:ratio}.}
$\Gamma_0\sim0$, which we assume here to be an equality. We see that the equation of
motion~\eqref{eq:motion_equation} is equivalent to the equation of a damped harmonic 
oscillator, where the Hubble parameter and decay rate act as a damping factor. 

It is known that considering the hilltop potential yields acceptable values of
only for a nonzero relative phase $\Theta$~\cite{Daido:2017wwb}, which is also
required for the axion scale $m_\Lambda$ to be nonvanishing. If $\Theta\neq 0$, 
the minimum of the hilltop term is shifted from $a_1=\pi f_a/\sqrt{3}$ and 
additional linear and cubic terms become relevant in the series approximation
of the potential.

Let $t_{\mathrm{osc}_m}$, $m=0,1$, denote the time at which the field $a_m$ begins to oscillate. 
The Universe is dominated by the oscillations of $a_1$ in the time interval between 
$t = t_{\mathrm{osc}_1}\simeq H^{-1}\sim M_\mathrm{Pl}/\sqrt{\rho({t}_{\mathrm{osc}_1})}$
and $t = \Gamma^{-1}_1$. The total energy density has contributions from three sources: 
the two densities $\rho_0$ and $\rho_1$ associated respectively with the fields $a_0$ and $a_1$,
and the radiation density due to the photon-$a_1$ coupling~\eqref{eq:coupling}. 
When $a_1$ starts to oscillate, the radiation energy density is close to zero, 
while the energy density of $a_1$ is much larger than that of $a_0$. 
This is based on the assumption that $m_\Lambda/M_c \gg 1$. Consequently, the 
hilltop term of the potential~\eqref{eq:rotated_potential} associated with $a_1$ 
has a greater contribution than the term proportional to $M^2_c$.
When the field $a_1$ begins to oscillate, its energy density is 
\begin{equation}
\rho_1(t_{\mathrm{osc}_1}) ~\simeq~ \Lambda^4\left( 1-\frac{\kappa}{ n^2} \right) 
-\frac{\Lambda^4}{f_a}\left[\sqrt{3}\,\Theta\,  a_1(t_{\mathrm{osc}_1})
+\frac{3(\kappa n^2-1) a_1(t_{\mathrm{osc}_1})^4}{8f^3_a}\right].
\end{equation}
This is a result of expanding the hilltop term in the potential, where the 
primary contributions are the linear, quartic and constant terms, and that the 
kinetic energy makes negligible contributions.
Thus, the start time of the oscillations of $a_1$ is
\begin{align}
    t_{\mathrm{osc}_1}~\simeq~ \frac{M_\mathrm{Pl}}{\sqrt{\rho_1(t_{\mathrm{osc}_1})}}
    ~\simeq~\frac{M_{\text{Pl}}}{\Lambda^4\sqrt{8/9-\sqrt{3}\Theta a_1(t_{\mathrm{osc}_1})/f_a-3 a_1(t_{\mathrm{osc}_1})^4/f^4_a}}\,,
\end{align}
where we have set $n=3$ and $\kappa = 1$.

\begin{figure}[t]
\centering
\includegraphics[width=\textwidth]{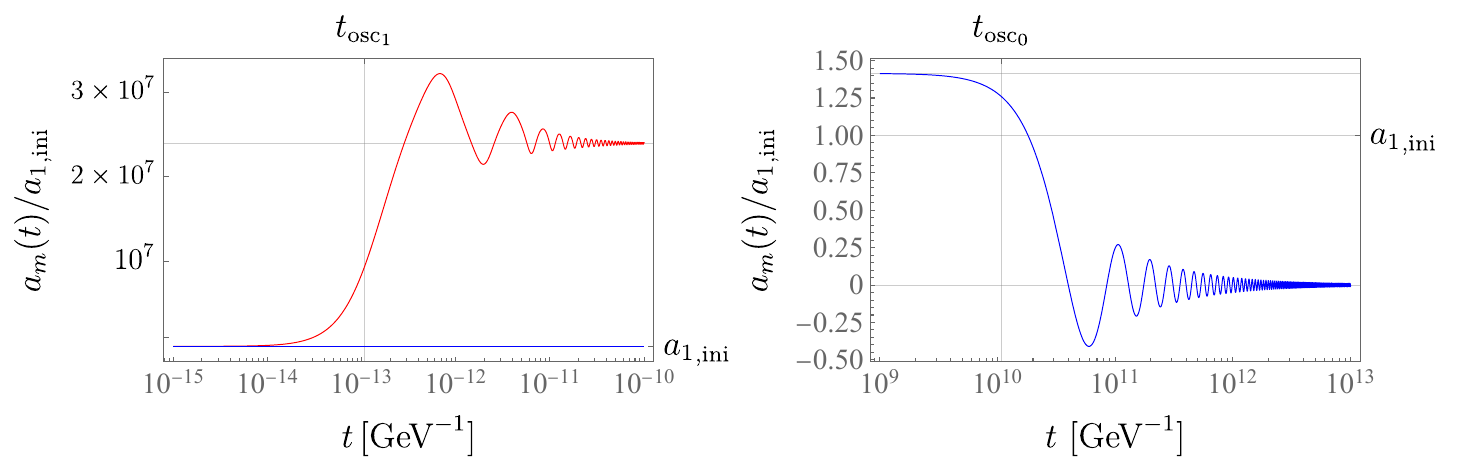}
\caption{Time evolution of the fields $a_0$ (blue) and $a_1$ (red), with initial 
         values $a_{1,\mathrm{ini}}=10^{12}/\sqrt{2}$\,GeV and $a_{0,\mathrm{ini}}=10^{12}$\,GeV. 
         The oscillations of $a_1$ begin at $t_{\mathrm{osc}_1}=1.06\times 10^{-13}\,\text{GeV}^{-1}$, 
         whereas the oscillations of $a_0$ start at $t_{\mathrm{osc}_0}=1.08\times 10^{10}\,\text{GeV}^{-1}$.
         \label{fig:fields_evolution}}
\end{figure}

On the other hand, the field $a_0$ evolves with the quadratic term. So, 
the following relation between the damping factor and the frequency 
determines its oscillation time
\begin{equation}
\label{eq:osc_condition}
    3H(t_{\mathrm{osc}_0}) ~=~ 4m_{a_0}^2\,.
\end{equation}
The above expression establishes a critical damping behavior of the 
oscillator, while for shorter times, we have an overdamped oscillator, 
and for longer times, we have an underdamped oscillator. 
For simplicity, we will assume that the following relation is satisfied
\begin{align}
\label{eq:ind_evol}
    t_{\mathrm{osc}_0}~\gg~ \Gamma^{-1}_1\,,
\end{align}
which implies that both fields evolve independently.
The decay rate of the field $a_1$ due to coupling with a photon pair
is given by
\begin{align}
    \Gamma_1~=~\frac{g^2_{a\gamma\gamma}{m}^3_{a_1}}{64 \, \pi}\,.
\end{align}
Since $a_0$ has a constant value at $t_{\mathrm{osc}_1}$, the 
hilltop term determines the evolution of the inflaton. 
Figure~\ref{fig:fields_evolution} illustrates the time evolution of $a_0$ and $a_1$,
with $a_{1,\mathrm{ini}}$ referring to the initial value of the ALP inflaton, 
set at $10^{12}/\sqrt{2}$\,GeV. The ratio of the initial values of the fields 
is given by eq.~\eqref{eq:initial_values}. The time at which the inflaton 
starts to oscillate is $t_{\mathrm{osc}_1}=1.06\times 10^{-13}\, \text{GeV}^{-1}$, while
the time at the onset of the oscillations of $a_0$ is 
$t_{\mathrm{osc}_0}=1.08\times 10^{10}\, \text{GeV}^{-1}$.

\section{Inflaxion}
\label{sec:Inflation}

We have established the structure of the time evolution of both the axion-like fields of
the simplest case of dark inflaxion. The first question we address is whether it is possible
that the inflationary period of our model complies with the constraints on the inflationary observables.

As mentioned earlier, the potential~\eqref{eq:rotated_potential} can serve as an inflationary model since 
the dynamics of the heavier field $a_1$ is governed by a hilltop term and a quadratic term,
both of which provide promising constructions for inflation. This may imply that the field $a_1$, 
with an adequate election of parameters, could reproduce a scenario similar to axion hilltop 
inflation~\cite{Daido:2017wwb}.
On the other hand, the mass of the field $a_0$ being much smaller than the mass of $a_1$, by 
construction, its dynamics start later than the inflationary dynamics. That is, the field 
$a_0$ is frozen during inflation, which means that the inflaton $a_1$ is the only dynamical 
field at this epoch.  As we will see in section~\ref{sec:densities}, $a_0$ can play the role
of a DM candidate.

Given the potential~\eqref{eq:rotated_potential} and considering only $a_1$ as the dynamical
degree of freedom during inflation, we can compute the slow-roll parameters
\begin{subequations}\label{eq:slow-roll4potential}
\begin{eqnarray}
\epsilon &=&  M^2_{\text{Pl}}\frac{3 n^2 \left[n \sin \left(\frac{\sqrt{3} a_1}{f_a}+\Theta \right)- \kappa  \sin \left(\frac{\sqrt{3} n a_1}{f_a}\right)+ \frac{f_an M^2_c}{3\sqrt{3}\Lambda^4 } (\sqrt{2}a_0-2a_1)\right]^2}{2f_a^2 \left[n^2  \cos \left(\frac{\sqrt{3} a_1}{f_a}+\Theta \right)- \kappa \cos \left(\frac{\sqrt{3} n a_1}{f_a}\right)+\frac{n^2M^2_c}{6\Lambda^4 } (a_0-\sqrt{2}a_1)^2\right]^2} \, , \label{eq:epsilon}\\   
\eta &=&M^2_{\text{Pl}}  \frac{3 \left[-\cos \left(\frac{\sqrt{3} a_1}{f_a}+\Theta \right)+\kappa \cos \left(\frac{\sqrt{3} n a_1}{f_a}\right)+\frac{2M^2_c}{9\Lambda^4 }\right]}{f^2_a \left[\cos \left(\frac{\sqrt{3} a_1}{f_a}+\Theta \right)-\frac{\kappa}{n^2} \cos \left(\frac{\sqrt{3} n a_1}{f_a}\right)+\frac{M^2_c}{6f^2_a\Lambda^4} (a_0-\sqrt{2}a_1)^2\right]}  \, . \label{eq:eta} 
\end{eqnarray}
\end{subequations}

As usual, while $\epsilon<1$ and $\eta<1$, $a_1$ could drive inflation. We define the 
end of inflation by imposing $\epsilon(a_{1,\mathrm{end}})\stackrel!= 1$.
The six parameters $\lbrace R, \Lambda, f_a, \Theta, \kappa, n\rbrace$ affect the 
slow-roll parameters~\eqref{eq:slow-roll4potential} and, consequently,
the predictions of the inflationary observables. 
If $M_c$ is sufficiently large (as in the case studied in section~\ref{subsubsec:BigMc}), 
then the potential behaves as a quadratic potential, which leads to 
the previously known chaotic inflation~\cite{Linde:1983gd}.
Chaotic inflation produces values of the tensor-to-scalar ratio $r\gtrsim\mathcal{O}\left(10^{-1}\right)$, 
see e.g.~\cite[Fig.~28]{Planck:2018vyg}, which are way too large compared with current
observations, eq.~\eqref{eq:r}. Hence, for inflation it is preferable that the hilltop term,
tuned by the size of $m_\Lambda$, dominates at the inflationary epoch. 

The parameters of our model can be constrained as follows.
We can fix the compactification radius (and hence $M_c$) by various considerations, such as 
those listed in table~\ref{table:limitsRadius}, but more importantly by DM constraints 
(see section~\ref{sec:densities}). This leads to 
$m_{a_0} = 1/\sqrt{3}R = 6.97\times 10^{-2}$\,eV.
As we saw in section~\ref{subsec:alps}, $\kappa$ and $n$ are related to the shape of the 
potential. $\kappa = 1$ produces a flat region near to the maximum of the potential, which 
is optimal for the inflationary constraint $\epsilon\ll 1$ to be fulfilled. Odd values
of $n$ result in a potential flat at the top and at the bottom. For these reasons,
we fix $\kappa=1$ and $n=3$, which are more convenient for an inflationary model.

\begin{figure}[t]
    \centering
    \includegraphics[scale=0.35]{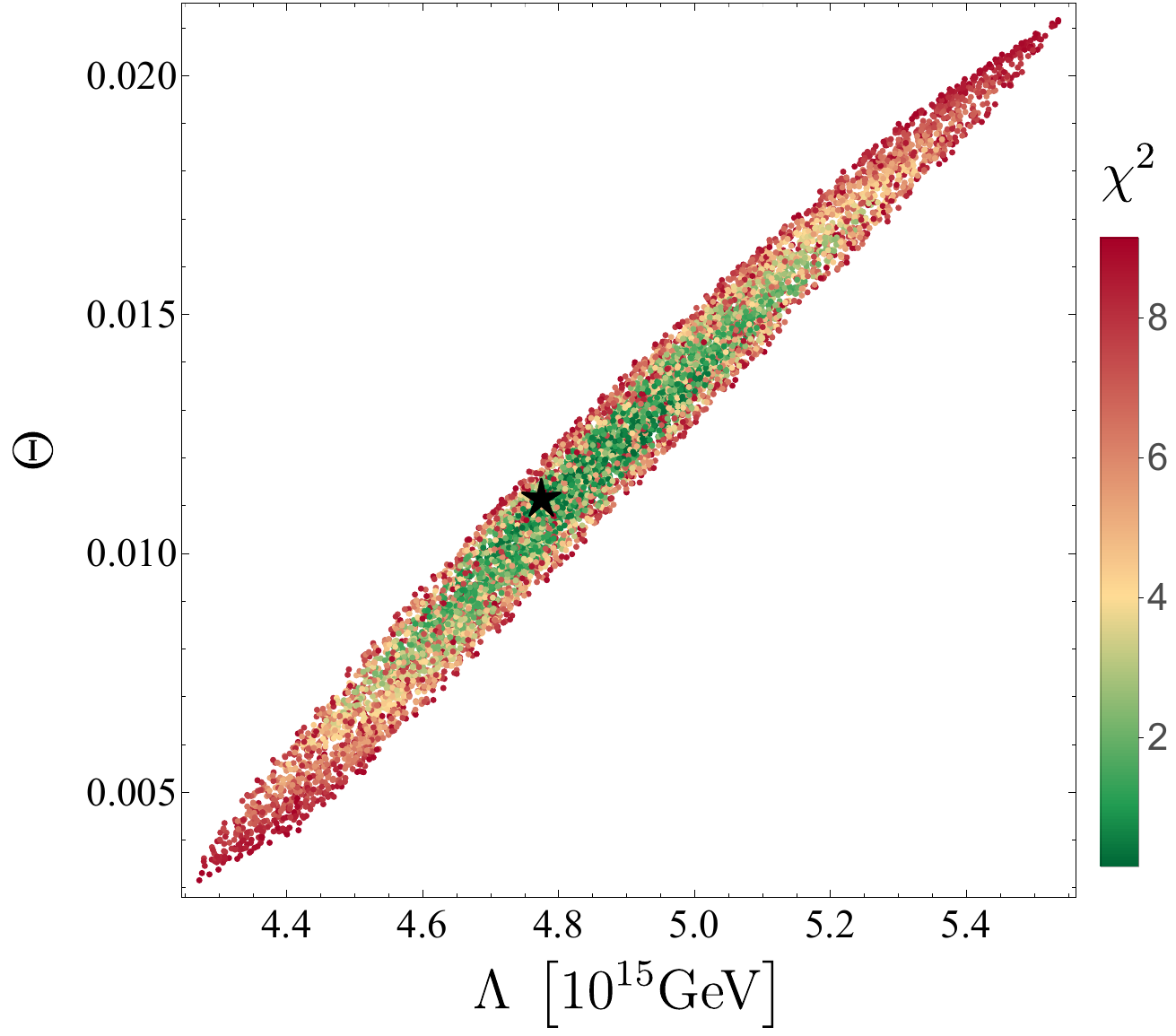}
    \caption{Values of $\Lambda$ and $\Theta$ with 
    $\chi_{n_s}^2+\chi^2_{A_s}+\chi^2_{\alpha_s}\leq 9$
    w.r.t.\ the observational values given by eqs.~\eqref{eq:n_s},~\eqref{eq:dnsdlnk} 
    and~\eqref{eq:As}. The number of e-folds is in the range $59 \leq N \leq 61$. 
    The star values correspond to the best fit $\chi^2=\chi_{n_s}^2+\chi^2_{A_s}+\chi^2_{\alpha_s}=0.154$. 
    All parameters associated with the best fit are given in table~\ref{tab:bestfit}.}
    \label{fig:parameters_region}
\end{figure}

The axion decay constant $f_a$ is restricted by the number of e-folds. 
$N$ does not reach values close to 60 if $f_a$ is not larger than
the reduced Planck mass. This is a property that this model shares with
axion hilltop inflation~\cite{Daido:2017wwb}, as we see in 
appendix~\ref{sec:AppendixA}.
We set $f_a=1.225\times10^{19}$\,GeV, so that inflation lasts long 
enough to solve the horizon and flatness problem. 

The phase $\Theta\neq 0$ is necessary to raise the scalar spectral index
$n_s$ to values close to the observation~\eqref{eq:n_s} and for $m_\Lambda\neq0$ 
(see eq.~\eqref{eq:m_lambda}). Further, the amplitude of scalar 
perturbations $A_s$ is sensible to the parameter $\Lambda$. 
These are the parameters we run in search of the best fit
of the model to the observations~\eqref{eq:Plancknsrvalues} and~\eqref{eq:PlanckAsdnsvalues}.

Given the inflationary observables, we find the best fit by minimizing $\chi^2$, which
determines the deviation of the theoretical values with respect to the observational data. 
We define it in the standard way as
\begin{equation}
\label{eq:chi2}
\chi^2 ~:=~ \chi^2_{n_s}+\chi^2_{r}+\chi^2_{A_s}+\chi^2_{\alpha_s}\, , 
\quad\text{where}\quad
\chi^2_z :=\frac{\left(z^\mathrm{th}  - z\right)^2}{\sigma_{z}^{2}}
\quad\text{and}\quad
z\in\lbrace n_s, r, A_s, \alpha_s\rbrace\,.
\end{equation}
Here, the superscript $\mathrm{th}$ denotes the theoretical prediction, while 
$\sigma_z$ corresponds to the measured error of $z$.

\begin{figure}[t]
    \centering
    \includegraphics[scale=0.35]{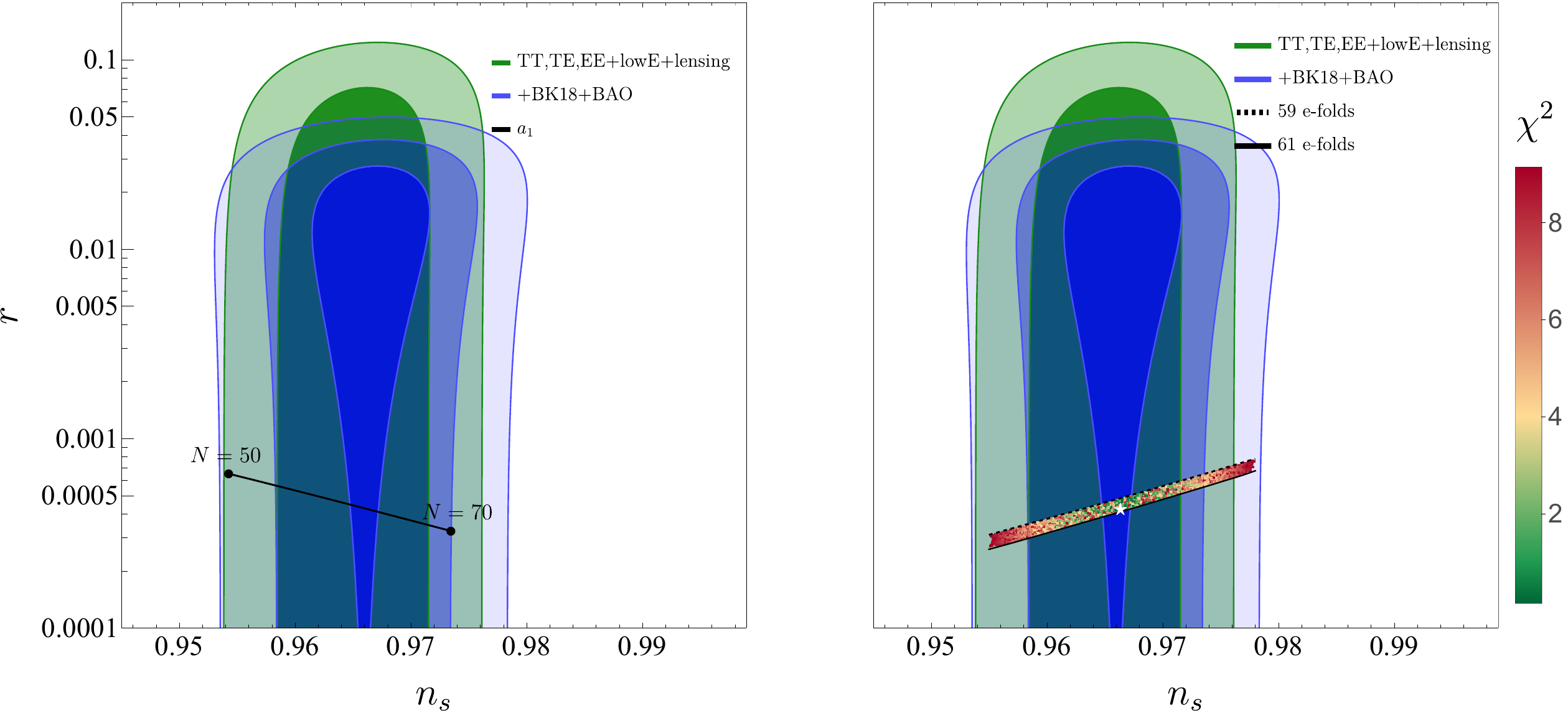}
    \caption{
    Predicted values for $n_s$ and $r$. The left panel shows the predictions
    for the best fit parameters (see table~\ref{tab:bestfit}) of the potential~\eqref{eq:rotated_potential} 
    running from $N=50$ to $N=70$ e-folds. The right panel displays the predictions
    for the parameters $\Lambda$ and $\Theta$ shown in figure~\ref{fig:parameters_region}.
    The green contours illustrate the $1\sigma$ and $2\sigma$ confidence level 
    regions of (TT,TE,EE+lowE+lensing) Planck's CMB data~\cite{Planck:2018vyg}. 
    The blue contours correspond to $1\sigma$, $2\sigma$ and $3\sigma$ confidence level regions 
    adding bounds from BAO and BICEP/Keck 18 (BK18) data~\cite{BICEP:2021xfz}.}
    \label{fig:ellipses}
   
\end{figure}

\begin{figure}[htpb!]
    \centering
    \includegraphics[scale=0.35]{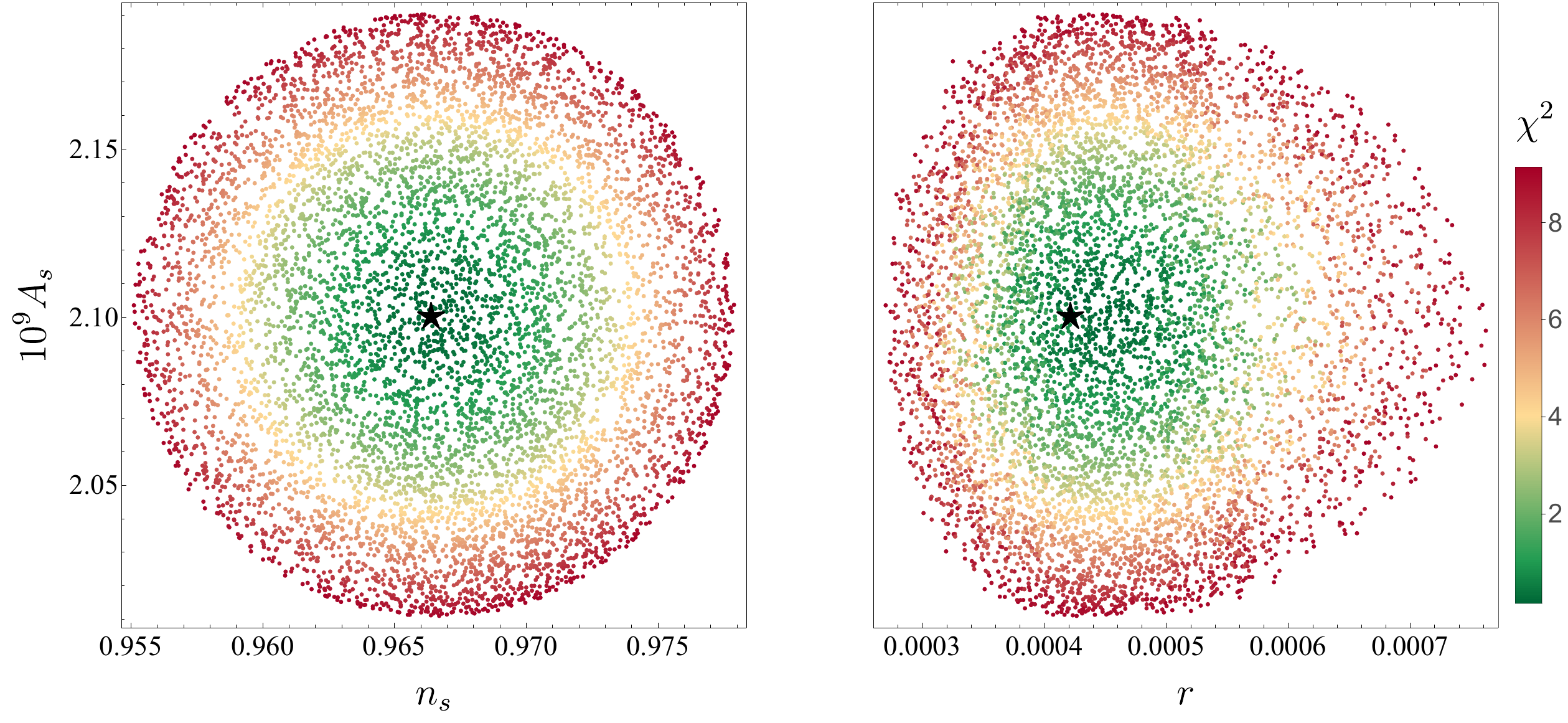}
    \caption{Predictions for $A_s$, $n_s$ and $r$ with the parameter values of 
    $\Lambda$ and $\Theta$ given in figure~\ref{fig:parameters_region}. 
    The star indicates the values that minimize $\chi_{n_s}^2+\chi^2_{A_s}+ 
    \chi^2_{\alpha_s}$.}
    \label{fig:two_density_plots}
\end{figure}

In figure~\ref{fig:parameters_region}, we show the region of the parameters $\Lambda$
and $\Theta$ with $\chi_{n_s}^2+\chi_{A_s}^2+\chi_{\alpha_s}^2 \leq 9$. We did 
not include the tensor-to-scalar ratio because we obtain the strong prediction 
$r~=~ \mathcal{O}\left(10^{-4}\right)$ for every choice of the parameters, which
is consistent with the reported bound $r<0.036$ at $95\%$ confidence from 
BK18~\cite{BICEP:2021xfz}. The star in the figure corresponds to the parameter 
values that minimize $\chi^2$.

We can now compute the values of $n_s$ and $r$ for different values of $N$. 
In figure~\ref{fig:ellipses}, we display the predicted values of these observables for 
$50\leq N \leq 70$. We used the parameters of the best fit given in table~\ref{tab:bestfit}. 
As we can see, all of these values are within the $3\sigma$ confidence level region 
from Planck and BK18. Note that, as stated before, $r=\mathcal{O}\left(10^{-4}\right)$ 
regardless of the number of e-folds. 

In figure~\ref{fig:ellipses}, we also show the predicted values of $n_s$ and $r$ for the
parameters $\Lambda$ and $\Theta$ indicated in figure~\ref{fig:parameters_region}. This 
region is bounded by the observed value of $n_s$ and the number of e-folds (black lines). 
Higher (lower) values of $r$ are associated with lower (higher) values of $N$.  
We note that the value of $r$ for the best fit is in the lower region, 
this is because of its number of e-folds ($N = 60.8$).
In figure~\ref{fig:two_density_plots}, we show the predicted values
of $A_s$, $n_s$ and $r$. The star depicts the best fit point
given in table~\ref{tab:bestfit} with
$\chi_{n_s}^2+\chi^2_{A_s}+\chi^2_{\alpha_s} = 0.154$.

In table~\ref{tab:bestfit}, we have included the parameter $\Psi$, which is
a global phase in the oscillatory components of the potential eq.~\eqref{eq:rotated_potential},
leading to
\begin{align}
\label{eq:additionalphase}
    V~\supset~\Lambda^4\left[ \cos\left(\frac{\sqrt{3}\,a_1}{f_a}+\Theta +\Psi \right) 
             -\frac{\kappa}{n^2}\cos\left(n\frac{\sqrt{3}\,a_1}{f_a}+ n\,\Psi\right)\right].
\end{align}
Note that this phase does not affect the  predictions of the inflationary observables. 
In particular, figures~\ref{fig:parameters_region},~\ref{fig:ellipses}, and~\ref{fig:two_density_plots} 
are identical for $\Psi = 0$ and $\Psi = 0.180$. The observables for our best
fit given in table \ref{tab:bestfit} are also the same. 
However, we use the phase $\Psi = 0.180$ because it shifts the value of the inflaton 
$a_1$ at the beginning and at the end of inflation, which shall be relevant
for the DM observables (see details in section~\ref{sec:densities}). 

\begin{table}[t]
\hspace{-.4cm}
    \parbox{.40\linewidth}{
        \centering
        \begin{tabular}{| c | c |}
            \hline
            parameter & value \\\hline
            \hline
            $\kappa$ & 1  \\  \hline
            $n$ & 3  \\ \hline
            $f_a$ & $1.225\times10^{19}$ GeV  \\ \hline
            $m_{a_0}$ & $6.97\times 10^{-2}$ eV  \\ \hline
            $\Lambda$ & $4.77\times 10^{15}$ GeV\\
            \hline
            $\Theta$ & 0.0111 \\ \hline
            $\Psi$ & 0.180 \\ \hline
            $g_{a\gamma\gamma}$ & $1.31\times 10^{-11}$ GeV$^{-1}$ \\ \hline 
            \multicolumn{2}{c}{} \\
        \end{tabular}
        }
    \hspace{.2cm}
    \parbox{.45\linewidth}{
        \centering
        \begin{tabular}{| c | c | c |}
            \hline
            observable & predicted value & observed value\\ \hline
            \hline
            $a_{1,\mathrm{ini}}$ & $7.07\times10^{11}$\,GeV  & --\\ \hline
            $a_{1,\mathrm{end}}$ & $8.14\times 10^{18}$\,GeV & -- \\ \hline
            $N$                  & 60.8                      & -- \\ \hline
            $n_s$                & 0.9664                    & $0.9665\pm0.0038$ \\ \hline
            $r$                  & $4.21\times 10^{-4}$      & $0.014^{+0.010}_{-0.011}$\\ \hline
            $10^9A_s$            & $2.100$                   & $2.105 \pm 0.030$\\ \hline
            $\alpha_s$           & $-9.15\times 10^{-4}$     & $-0.006\pm 0.013$\\ \hline
            $\chi_{n_s}^2+\chi^2_{A_s}$ & $8.21 \times 10^{-4}$     & --\\\hline
            $\chi_{n_s}^2+\chi^2_{A_s}+\chi^2_{\alpha_s}$ & $0.154$ & -- \\ \hline
         \end{tabular}
       }
       \caption{Parameter values that minimize $\chi_{n_s}^2+\chi^2_{A_s}+\chi^2_{\alpha_s}$ 
                contrasted with the inflationary observational values given 
                in eqs.~\eqref{eq:Plancknsrvalues} and~\eqref{eq:PlanckAsdnsvalues}.
                \label{tab:bestfit}}
\end{table}

Note that we obtained a super-Planckian field range 
$\Delta a_1= a_{1,\mathrm{end}} - a_{1,\mathrm{ini}}= 3.34\,M_\mathrm{Pl}$, which automatically
satisfies the Lyth bound~\cite{Lyth:1996im} 
\begin{equation}
\label{eq:lyth}
\frac{\Delta a_1}{M_\mathrm{Pl}} ~\geq~ 2\times \left(\frac{r}{0.01} \right)^{\nicefrac12}\simeq 2.37\, .
\end{equation} 
  
In contrast to other results~\cite{Daido:2017wwb}, our predicted values of the Hubble parameter 
are large, $3.9\times 10^{12}\,\mathrm{GeV}\lesssim H_*\lesssim 7.1\times 10^{12}\,\mathrm{GeV}$, 
with a value of $H_*\simeq 5.1\times 10^{12}$\,GeV for our best fit. The total number 
of e-folds is between $59<N<61$, large enough to explain the flatness and horizon problem. 
Further, our predictions establish that inflation must have lasted about
$5.5\times 10^{-36}\,\text{s}<t_\mathrm{inf}<1.0\times 10^{-35}\,\text{s}$,
and $t_\mathrm{inf}\simeq 7.9\times 10^{-36}\,\text{s}$ at the best fit point. 
These results are closer to the expected inflation duration value  
$t_\mathrm{inf} \lesssim 10^{-30}$\,s~\cite{Kolb:1990vq} 
than the predictions in typical axion hilltop inflation (see appendix~\ref{sec:AppendixA}).

\section{DM dynamics after inflaxion}
\label{sec:densities}

The axion-like inflaton $a_1$ must evolve until the end of inflation, when
it must decay to radiation, leading to a radiation-dominated epoch. After
some time, $a_0$ should start its oscillating dynamics and diluting with an equation of state 
$\omega \simeq 0$, leading the matter-dominated epoch. Due to its lacking
of couplings to other fields, it becomes an interesting DM candidate
whose abundance must comply with the observation of $\Omega_\mathrm{DM}$.
Let us study the details of this process by following the evolution of
the energy densities as the Universe expands.

%%%%%%%%%%%%%%%%%%%%%%%%%%%%%%%%%%%%%%
\subsection{Matter and radiation}

The misalignment mechanism assumes that the field in the early Universe, prior
to the onset of the nontrivial potential that governs inflation, possesses 
an arbitrary initial value. This value approximates the value of the field 
when $3H(t)$ displays critical damping behavior.
The criterion to distinguish between matter and radiation 
energy densities is their behavior as the expansion of the Universe dilutes them.
An energy density is classified as matter energy density if it decreases as 
$\mathrm{a}^{-3}$, and it is considered radiation energy density if it 
falls off as $\mathrm{a}^{-4}$.

\begin{figure}[t]
\centering
\includegraphics[width=\textwidth]{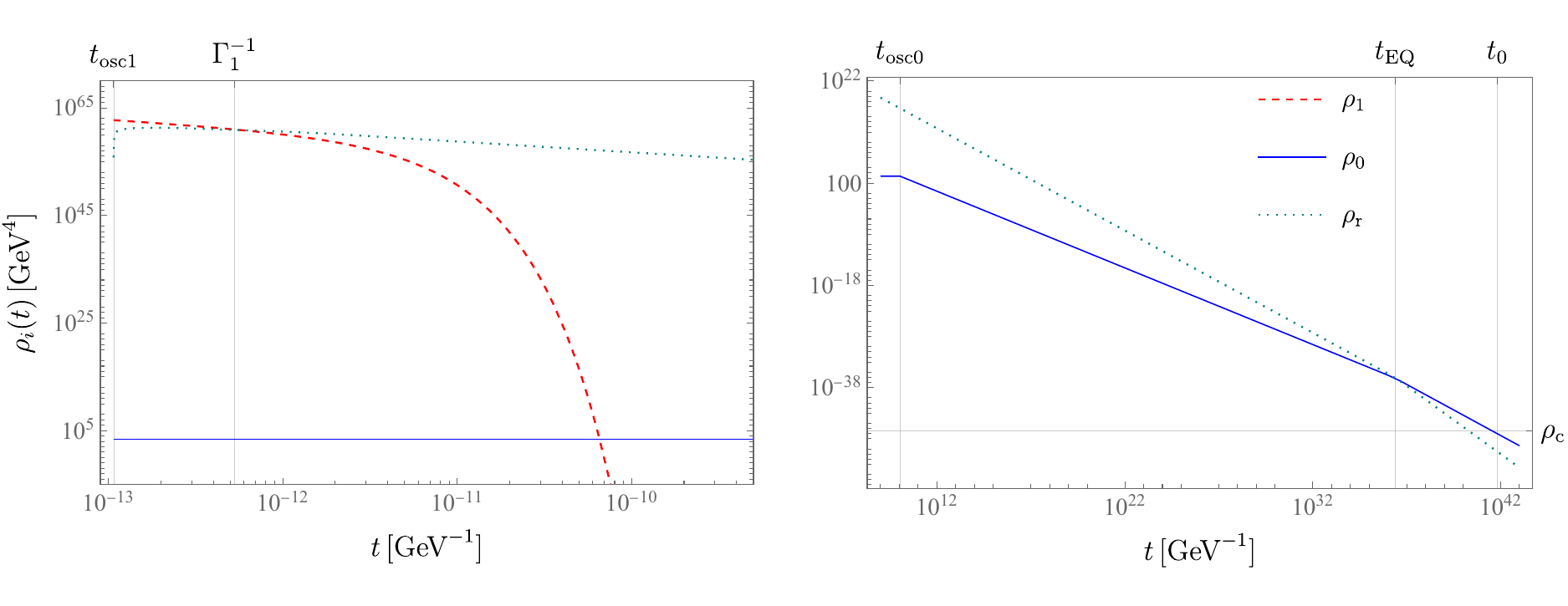}
\caption{Time evolution of $\rho_1$, $\rho_0$ and $\rho_\mathrm{r}$.
         The energy density of the field $a_1$ is represented by the dashed line, while the 
         solid line corresponds to the $a_0$ energy density, and the dotted line denotes $\rho_\mathrm{r}$. 
         The times shown are $t_{\mathrm{osc}_1}=1.06\times 10^{-13}$\,GeV$^{-1}$, 
         $t_{\mathrm{osc}_0}=1.08\times 10^{10}$\,GeV$^{-1}$, 
         $\Gamma_1^{-1}=5.28\times 10^{-13}$\,GeV$^{-1}$, 
         $t_{\mathrm{EQ}}\simeq 2.44\times 10^{36}$\,GeV$^{-1}$, 
         $t_0\simeq 6.53\times 10^{41}$\,GeV$^{-1}$. $\rho_\mathrm{c}$ denotes 
         today's critical energy density.
         \label{fig:densityEvolution}}
\end{figure}

The time evolution of the energy densities of $a_0$ and $a_1$ along with
the radiation energy density $\rho_\mathrm{r}$ is described by the equations
\begin{subequations}
   \label{eq:ED_energy_densities}
   \begin{eqnarray}
            \dot{\rho}_1+[3H(t)+\Gamma_1](\omega+1)\rho_1 &=&0,\\
   \label{eq:sol_rho_0}
	    \dot{\rho}_0+3H(t)\rho_0&=&0,\\     
	    \dot{\rho}_\mathrm{r}+4H(t)\rho_\mathrm{r}&=&(\omega + 1)\,\Gamma_1\,  \rho_1,
   \end{eqnarray} 
\end{subequations}
where $\omega$ is given by the equation of state. 
The energy densities of $a_0$ and $a_1$ behave as vacuum energy densities
before the fields oscillate, as they stay at their initial values $a_{0,\mathrm{ini}}$
$a_{1,\mathrm{ini}}$, as discussed in section~\ref{sec:dynamics}, see figure~\ref{fig:fields_evolution}.
Provided that the fields stay at their initial constant values, their equation
of state is just $\omega=V(a_{m,\mathrm{ini}})/(-V(a_{m,\mathrm{ini}})) =-1$, $m=0,1$.
The significant evolution of energy densities occurs from the time $t_{\mathrm{osc}_m}$ 
when the fields start their oscillation around the minimum of their potential. 
The solutions to eqs.~\eqref{eq:ED_energy_densities} are
\begin{subequations}
   \begin{align}
\label{eq:den_rho}
      \rho_0(t)~=~& \, \rho_0(t_{\mathrm{osc}_0})\left( \frac{\mathrm{a}
      (t_{\mathrm{osc}_0})}{\mathrm{a}(t)}\right)^3 ,\\
      \rho_1(t)~=~& \,\rho_1(t_{\mathrm{osc}_1})\left( \frac{\mathrm{a}
      (t_{\mathrm{osc}_1})}{\mathrm{a}(t)}\right)^{3(\omega + 1)}e^{-
      \Gamma_1(t-t_{\mathrm{osc}_1})},\\
      \rho_\mathrm{r}(t)~=~& \, \mathrm{a}(t)^{-4}\left[\rho_1(t_{\mathrm{osc}_1})\mathrm{a}
      (t_{\mathrm{osc}_1})^{3(\omega+1)}I(t)\right],
   \end{align} 
\end{subequations}
where we defined
\begin{equation}
I(t)~:=~(\omega + 1)\Gamma_1\int^{t}_{t_{\mathrm{osc}_1}}\mathrm{d}t'
        e^{-(\omega + 1)\Gamma_1 (t'-t_{\mathrm{osc}_1})} \mathrm{a}(t')^{4-3(\omega + 1)}\,.
\end{equation}
Figure~\ref{fig:densityEvolution} displays the behavior of all energy densities 
from the onset of the inflaton dynamics $t_{\mathrm{osc}_1}$ until today $t_0$.
\begin{figure}
\centering
  \includegraphics[scale=0.7]{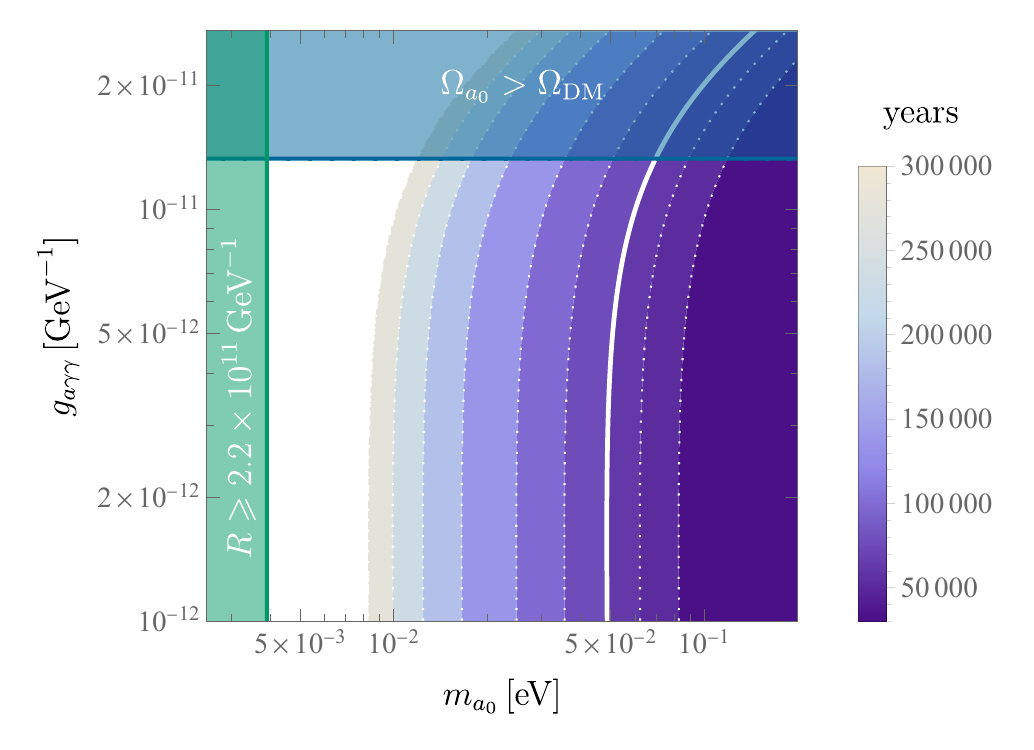}  
\caption{Time of matter-radiation equality $t_{\mathrm{EQ}}$ as a function of $g_{a\gamma\gamma}$
        and the mass of $a_0$. The white curve corresponds to the usual expectation of 
        $t_{\mathrm{EQ}}=51, 100$\,years. The green region is excluded by the constraints 
        on the compactification radius shown in table~\ref{table:limitsRadius}. 
        The upper blue region is excluded by the measured value of $\Omega_\mathrm{DM}$, 
        see eq.~\eqref{eq:OmegaDMtoday} and figure~\ref{fig:abundance}. 
        \label{fig:equality_time}}
\end{figure}
\begin{figure}[t]
\centering
\includegraphics[scale=.83]{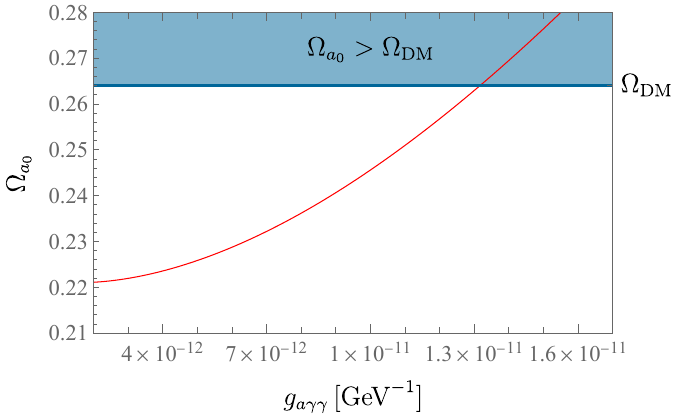}
    \caption{Abundance of the field $a_0$, as a function of the coupling $g_{a\gamma\gamma}$. 
             The horizontal line denotes the abundance of cold DM reported by 
             (TT,TE,EE+lowE+lensing) Planck's CMB data~\cite{Planck:2018vyg}. 
             The blue region is excluded since $\Omega_{a_0}$ must not exceed 
             $\Omega_\mathrm{DM}$.
             \label{fig:abundance}}
\end{figure}
From eq.~\eqref{eq:den_rho} we observe that the energy 
density of the field $a_0$ is diluted as matter energy density. This fact along with 
its lack of coupling to photons or other fields, let us identify the field $a_0$ 
as DM.

The time evolution of DM energy density depends on the type of energy density 
that dominates the Universe at each time. If radiation prevails,
the Hubble parameter equals $1/2t$; in contrast, if matter dominates, the 
Hubble parameter equals $2/3t$. The explicit time dependence of the DM 
energy density in the two scenarios is then
\begin{align}
     \rho_0(t) ~=~ \rho_0(t_{\mathrm{osc}_0})t^{3/2}_{\mathrm{osc}_0}\times
     \begin{cases} t^{-3/2}                     &\qquad t_{\mathrm{osc}_0}\lesssim t \lesssim t_\mathrm{EQ},\\
                   t^{-2}t^{1/2}_{\mathrm{EQ}}  &\qquad t_\mathrm{EQ} \lesssim t,
     \end{cases}
\end{align}
where $t_\mathrm{EQ}$ denotes the time of matter-radiation equality.
Assuming that the energy density of baryonic matter has no substantial
impact on the matter density at the time of equality between radiation 
and matter, one can obtain $t_\mathrm{EQ}$ from the equation
\begin{align}
\label{eq:teq}
    \rho_0(t_\mathrm{EQ})~=~\rho_\mathrm{r}(t_\mathrm{EQ})\,.
\end{align}
The solution to this equation depends particularly on the mass $m_{a_0}$ of the 
DM candidate $a_0$ and the coupling $g_{a\gamma\gamma}$. Figure~\ref{fig:equality_time}
shows the values of $t_\mathrm{EQ}$ in such parameter space. The white curve corresponds
to the values of $m_{a_0}$ and $g_{a\gamma\gamma}$ that best fit the expectation of 
$t_\mathrm{EQ}=51,100$\,years~\cite{Workman:2022ynf}. We also present the regions
that are excluded by i) the bounds on the compactification radius, as presented
in table~\ref{table:limitsRadius}, since it determines $m_{a_0}$, and ii) the current 
abundance of DM as we now discuss.

We can compute the abundance of $a_0$ and compare it
with the observations of the CMB. The abundance of the field $a_0$ is defined by
\begin{equation}
\label{eq:abundance}
\Omega_{a_0} = \frac{\rho_0(t_0)}{\rho_\mathrm{c}} \, , 
\end{equation}
where $t_0$ is the current time and $\rho_\mathrm{c} := 3H^2/8\pi G$ is the critical density today.
The abundance of cold DM reported by Planck is
\begin{equation}
\label{eq:OmegaDMtoday}
\Omega_\text{DM} h^2 =  0.11933 \pm 0.00091 , \,\, \text{({\small TT,TE,EE+lowE+lensing+BAO,~\cite[Table 2]{Planck:2018vyg}})}\, ,
\end{equation}
where $h = 0.6766 \pm 0.0042$. In the figure~\ref{fig:abundance}, we show
the predicted values of $\Omega_\text{DM}$ for different values
of the coupling $g_{a\gamma\gamma}$. 

The relation between the energy density $\rho_0(t)$ and the initial value of the 
field $a_0$ is established in $\rho_0(t_{\mathrm{osc}_0})$, where $t_{\mathrm{osc}_0}$ 
corresponds to the instant when the field starts oscillating. This initial value 
dependence is inherited in the abundance of $a_0$, shown in eq.~\eqref{eq:abundance}.
Nevertheless, eq.~\eqref{eq:initial_values} reveals that the initial values of 
$a_1$ and $a_0$ are not independent. Additionally, the initial value of the $a_1$ 
field is determined through inflation, as explored in section~\ref{sec:Inflation}.
If the initial value of the inflaton is of the order $\mathcal{O}(M_\mathrm{Pl})$, 
due to the relationship mentioned above,
the mass required by $a_0$ to achieve acceptable equilibrium time and DM abundance
values is excluded by constraints on the compactification radius. 
To be explicit, the mass would be in this case of the order of $10^{-36}$\,GeV and 
the compactification radius is of the order of $10^{20}$\,m. In order to avoid this, we introduce a global
phase $\Psi$ into the hilltop term of the potential~\eqref{eq:rotated_potential},
taking the form given by eq.~\eqref{eq:additionalphase}, to translate the initial value of the
inflaton, as well as the initial value of $a_0$. It is important to note that the shape 
of the potential remains unchanged regardless of the value of $\Psi$, since it is a global 
phase. The predictions for the inflationary parameters, including $n_s$, $r$, $N$, $\alpha_s$,
and $A_s$, remain unaffected. Once the global phase is introduced,
the initial value of the $a_0$ field may be shifted to values approximately equal to 
$10^{12}$\,GeV, thus enabling us to obtain $m_{a_0}\approx10^{-11}$\,GeV, which 
ensures the consistency with $\Omega_\mathrm{DM}$ and $t_\mathrm{EQ}$. This mass value
also satisfies the constraints imposed on the compactification radius. The global 
phase for this scenario is chosen to be $\Psi=0.180$.

We find that the best fit value of the axion-photon coupling is 
$g_{a\gamma\gamma} = 1.31\times 10^{-11}$\,GeV$^{-1}$, corresponding
to the total abundance of cold DM. Values greater than this
are excluded. Therefore, the field $a_0$ not only has an energy density that
behaves as matter but also can reproduce the abundance of DM today,
completing so the description of inflation and DM of dark inflaxion.

\section{Final remarks and outlook}
\label{sec:conclusions}

When an extra dimension is compactified, the dynamics of a 5-d ALP in a hilltop 
potential induces 4-d KK axion-like fields with identical decay constant that can 
behave as inflaton(s) or DM. As explained in section~\ref{sec:benchmark}, the effective 4-d 
potential receives, in addition to the hilltop structure, a quadratic contribution 
for all 4-d modes, which is inherited from the 5-d ALP kinetic energy. Considering
the full 4-d action, this model exhibits various behaviors depending on the values
of the compactification scale $M_c=1/R$ and the natural axion scale $m_\Lambda$ (associated
with the quotient $\Lambda^2/f_a$). Assuming that a subset of the heavier fields 
are decoupled from low-energy observable physics due to their huge masses, we find 
three interesting possibilities of effective cosmological physics:
\begin{enumerate}
\item[i)]   If $m_\Lambda\gg M_c$, the heaviest undecoupled axion is much heavier than the 
rest, which have masses around $M_c$. In this case, the heaviest field drives 
hilltop inflation, while the lighter fields act like DM.
\item[ii)]  When $m_\Lambda \approx M_c$, all axions have a mass of the same order. This
case is not compatible with constraints on the reheating temperature after inflation, but 
could very well be associated with a description of DM (probably dynamical dark matter).
\item[iii)] For $m_\Lambda\ll M_c$, we have a very light axion, while the others have 
approximately the same mass. One conceives a multifield inflationary scenario at a 
scale around $M_c$, whose foreseeably complicated properties will be studied elsewhere.
\end{enumerate}
Motivated by the mixed nature of the physics emerging from a single 5-d 
field, we call our model {\it dark inflaxion}.

In the simplest scenario of the case i) of dark inflaxion, one identifies the inflaton 
to be the hilltop direction in field space, which is a mass eigenstate and corresponds 
to the only dynamical field at earlier times. As the expansion of the Universe proceeds, 
all lighter fields begin successively their oscillatory matter dynamics, from the 
heavier to the lightest one.

We have illustrated this scenario considering that only the dynamics of two 
4-d KK modes is relevant at low energies while $m_\Lambda\gg M_c$. The heavier
field plays the role of the inflaton, while the lighter behaves as stable DM. The parameters
of the effective hilltop inflationary model can be fitted to accommodate the 
inflationary observables $n_s,r,A_s$ and $\alpha_s$ with $\chi^2\simeq0.154$,
as shown in figures~\ref{fig:ellipses} and~\ref{fig:two_density_plots}, and 
table~\ref{tab:bestfit}, including also the right timing and duration of inflation.
In section~\ref{sec:densities}, we have computed the dynamics of the energy 
densities of the matter fields and light, where it becomes explicit that
the lighter field behaves as matter. Further, it is possible to constrain 
the coupling of the 4-d ALPs to comply with the observed DM abundance and
retain compatibility with the expected time of matter-radiation equality
and enough amount of reheating, 
while avoiding the known constraints on the compactification scale.

Our current study shows the promising features of dark inflaxion, which appears
naturally in models with extra dimensions, including those arising from
string compactifications. However, there are some extant questions. We still
need to find a dynamic mechanism to fix the hierarchy $m_\Lambda\gg M_c$ chosen 
in this work. This seems to be challenging without invoking additional
interactions. Instead, one might be interested in addressing the more general
case with intertwined dynamics. In particular, one should explore the 
possibilities and consequences that yield multifield inflation and the 
details of the effective DDM case. This work will be done elsewhere.

\section*{Acknowledgments}
It is a pleasure to thank J.~A.~Cort\'es Asencio for insightful discussions at early
stages of this project, and O.~P\'erez-Figueroa for enlightening suggestions. This 
work was supported by UNAM-PAPIIT IN113223, CONACyT grant CB-2017-2018/A1-S-13051,
and Marcos Moshinsky Foundation.

\appendix

\newpage

\section{Decay constant in the ALP miracle}\label{sec:AppendixA}

\begin{figure}[b]
\centering
\includegraphics[scale=0.38]{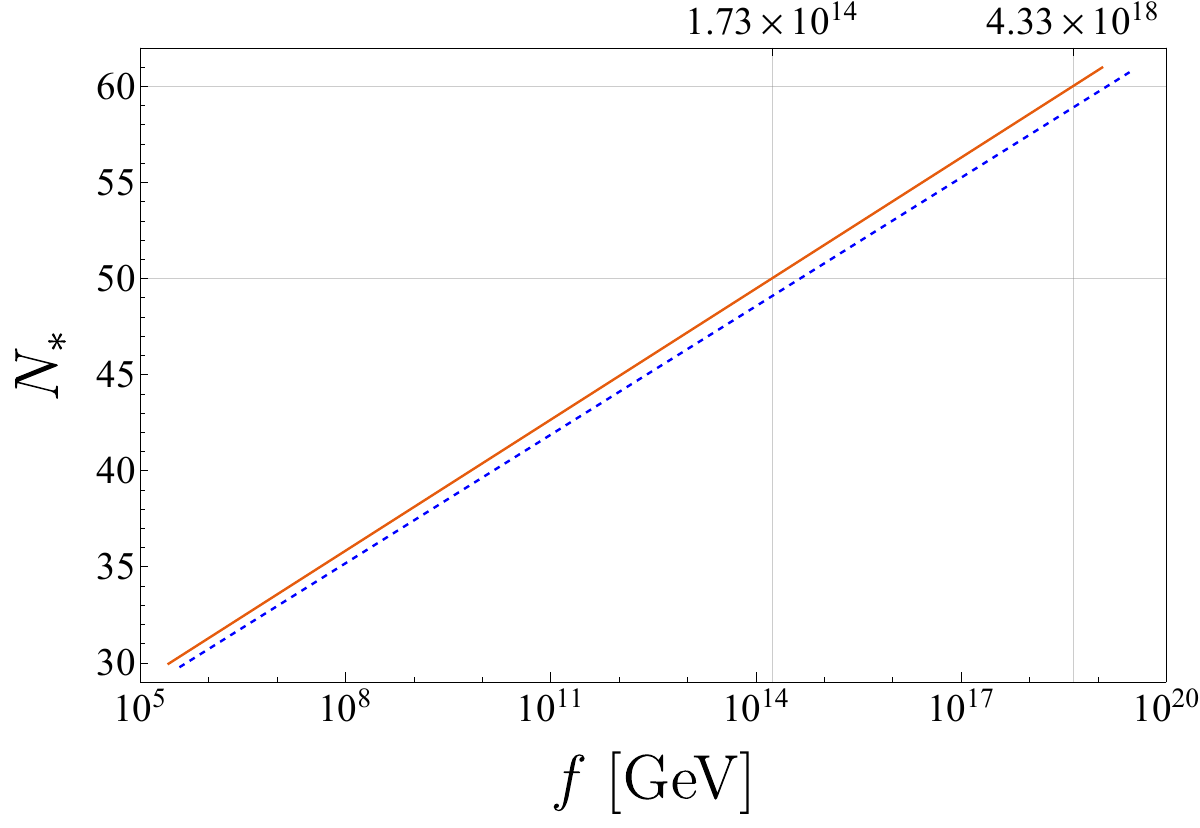}
\caption{Number of e-folds $N_*$ as a function of the decay constant $f$
from the potential~\eqref{eq:hilltop_potential}. We show solution 1 (solid) and
solution 2 (dashed).
\label{fig:NvsF}}
\end{figure}

In the ALP miracle~\cite{Daido:2017wwb}, one adopts the hilltop 
potential~\eqref{eq:hilltop_potential} for the inflaton. 
This can be approximated by a quartic potential close to the minimum, 
so the energy density of the inflaton evolves like radiation once inflation ends. 
The number of e-folds is given by
\begin{equation}\label{eq:Nefoldsalpmiracle}
N_*~\simeq~ 
\int_{\phi_{\text{end}}}^{\phi_{*}}\frac{3H^2 }{V'}d\phi ~\simeq~ 61 + \ln \left( \frac{H_*}{H_\text{end}}\right)^{1/2}  + \ln \left(\frac{H_\text{end}}{10^{14}\,\text{GeV}}\right)^{1/2}\, ,  
\end{equation}
where the subscript $*$ indicates that the variable is evaluated
when the cosmological scale leaves the horizon. 
The Hubble parameter is almost constant during inflation,
so it is natural to set $H_*\simeq H_\text{end}$.
In the original reference two interesting values of $\Theta$ associated 
with optimal values of the scalar spectral index $n_s$ are studied. They
are referred to as {\it solution 1 (2)} for a larger (smaller) value
of $\Theta$. In these cases, the decay constant is related to the Hubble parameter by
\begin{subequations}
\begin{align}
f &\simeq 9.70\times 10^{11}\, \text{GeV}\,\left(\frac{n}{3}\right)^\frac{1}{2}\left(\frac{H_*}{1\, \text{GeV}}\right)^{0.506}\;\;\;\; \text{for solution 1}\,  ,\\
f &\simeq 1.83\times 10^{12}\, \text{GeV}\,\left(\frac{n}{3}\right)^\frac{1}{2} \left(\frac{H_*}{1\, \text{GeV}}\right)^{0.514}\;\;\;\; \text{for solution 2}\,  . 
\end{align}
\end{subequations}

In~\cite{Daido:2017wwb} $10^{-12}\,\text{GeV}< H_* < 10^{-8}\,\text{GeV}$ is reported, 
corresponding to a decay constant
$8.11\times 10^{5}\,\text{GeV}<f<8.61\times 10^7\,\text{GeV}$ for solution 1
and $1.45\times 10^{6}\,\text{GeV}<f<1.68\times 10^8\,\text{GeV}$ for solution 2. 
Interestingly, these values of $H_*$ (and $f$) seem to be too low for a consistent 
model of cosmic inflation, as one can see by studying e.g.\ the number of inflationary 
e-folds and the duration of inflation. We can use eq.~\eqref{eq:Nefoldsalpmiracle} to 
compute this number, using that $H_*\simeq H_\text{end}$. We find that the consequent
e-folding number is $30  \lesssim N_*  \lesssim 36$, which is not enough to solve the 
horizon and flatness problem.  We plot the naive relation between the axion decay 
constant and the number of e-folds in figure~\ref{fig:NvsF}. 
One finds that, in principle, $N_*$ cannot reach the expected value of $60$ e-folds 
if the decay constant is sub-Planckian. On the other hand, the simplest approach to
compute the duration of inflation yields $t_\mathrm{inf}\simeq N_*/H_*$. Using the 
mentioned values of $N_*$ and $H_*$, we find
$2.4\times 10^{-15}\,\text{s}\lesssim \Delta t_\text{inf} \lesssim 
2\times 10^{-11}\,\text{s}$. In the most natural models, one expects
that inflation happens and lasts about $10^{-30}$\,s~\cite{Kolb:1990vq}. This suggests
that in inflationary models based on ALPs with sub-Planckian axion decay constant, one
must still explain whether these small numbers of e-folds and large duration times 
are adequate to explain cosmic inflation or they should be amended.

%%%%%%%%%%%%%%%%%%%%%%%%%%%%%%%%%%%%%%%%%%%%%%%%%%%%%%%%%%%%%%%%%%%%%%%%%%%%%%%%%%%%%%%%%%%%%%%%%%%%%%%%%%%%%%

%%%%%%%%%%%%%%%%%%%%%%%%%%%%%%%%%%%%%%%%%%%%%%%%%%%%%%%%%%%%%%%%%%%%%%%%%%%%%%%%%%%%%%%%%%%%%%%%%%%%%%%%%%%%%%
%\bibliographystyle{OurBibTeX}
%\bibliography{DM}

\providecommand{\bysame}{\leavevmode\hbox to3em{\hrulefill}\thinspace}

\end{document}